\newif\ifSC
\SCtrue
\ifSC
\documentclass[onecolumn,draftclsnofoot,12pt]{IEEEtran}
\else
\documentclass[twocolumn,10pt]{IEEEtran}
\fi
\usepackage{etoolbox}
\usepackage{subfigure}

\newtoggle{SC}
\togglefalse{SC}
\ifSC
\toggletrue{SC}
\fi

\usepackage{bbm}
\usepackage{amssymb}
\usepackage{amsmath,amsthm}
\usepackage{color}
\usepackage{amsfonts}
\usepackage{graphicx}
\usepackage{verbatim}
\usepackage{epstopdf}
\usepackage{cite}
\usepackage{tabulary}
\usepackage{mathtools}

\newcommand\expect[1]{\mathbb{E}\left[#1\right]}
\newcommand\prob[1]{\mathbb{P}\left[#1\right]}

\newcommand\indside[1]{\mathbbm{1}\left({#1}\right)}

\newcommand{\expects}[2]{\mathbb{E}_{#1}\left[#2\right] }

\newcommand{\event}[1]{\mathcal{#1}}
\newcommand{\point}[1]{\mathbf{#1}}

\newcommand{\dd}{\mathrm{d}}

\newcommand{\expU}[1]{e^{#1}}

\newcommand{\beq}{\gamma}
\newcommand{\Q}{W}
\newcommand{\s}{\mathcal{N}}

\newcommand{\ths}{\text{th}}
\newcommand{\pR}{p_\mathrm{R}}

\newcommand{\Z}{Z}

\newcommand{\LBI}[1]{\underline{\underline{#1}}}
\newcommand{\LBII}[1]{\underline{#1}}

\newcommand{\DEP}{\mathrm{D}}

\newcommand{\pRIND}{\overline{p_\mathrm{R}}}
\newcommand{\pRDEP}{\underline{p_\mathrm{R}}}

\newtheorem{theorem}{Theorem}
\newtheorem{lemma}{Lemma}
\newtheorem{corollary}{Corollary}

\newcommand{\Lmax}{L_\mathrm{max}}

\newcommand{\sbangle}{\omega}
\newcommand{\lbuild}{\mu}
\newcommand{\lbs}{\lambda}
\newcommand{\PBuild}{\psi}

\newcommand{\omicron}{\mathrm{o}}

\newcommand{\PBS}{\Lambda}

\iftoggle{SC}{}
{}
\iftoggle{SC}{}
{}

\iftoggle{SC}{\newcommand{\begs}{\begin{small}}}{\newcommand{\begs}{}}
\iftoggle{SC}{\newcommand{\ens}{\end{small}}}{\newcommand{\ens}{}}

\newcommand{\insertnotationtable}{
\begin{table}[t!]\label{notation}
	\caption{Summary of Notation}
	\begin{tabulary}{\columnwidth}{ |l | L | }\hline
		{\bf Notation} &{\bf Description}\\ \hline
		$\PBS,\lbs$ & The PPP modeling the BS locations, its  density. \\ \hline
		$\Psi,\PBuild,\lbuild$ & The Boolean model for the blockage locations, the PPP modeling the centers of blockages and its density. \\ \hline
		$\ell_k,\theta_k$ & The length and orientation of the $k\ths$ blockage. \\ \hline
		$F_L(),F_\Theta()$ & The distribution of blockage lengths $\ell$ and orientations $\theta$. $F_L(\event{L})$ denotes the probability that $\ell\in\event{L}$.\\ \hline
			$\Psi(\ell,\theta)$ & The derived blockage process from $\Psi$ consisting of blockages having length between $\ell$ and $\ell+\dd\ell$ and orientation between $\theta$ and $\theta+\dd\theta$.  \\ \hline
			$\beta,L_m$ & A parameter for the blockage process defined as $\lbuild \expect{\ell}\frac2\pi$, the maximum blockage length.  \\ \hline
		$\point{B}_j,R_i,\point{\Z}_j$ & The $j\ths$ closest BS from the user at the origin, the distance of this BS from the user, the link between this BS and the origin. \\ \hline
		$n$ & The order of macro diversity {i.e.} the number of simultaneous connected BSs.\\\hline
		$\Phi_i$ & The angle between $\point{Z}_j$ and $\point{Z}_n$. \\ \hline
		$\Phi$ & The angle between $\point{Z}_1$ and $\point{Z}_2$ for  special case of second order diversity.\\ \hline
		$\mathcal{A}_j$ & The event  that the $j\ths$ link is LOS.  \\ \hline
		  $\pR$ & The probability of reliability.  \\ \hline
		 $\mathcal{P}_i$ & The parallelogram constructed on the $i\ths$ link as shown in Fig. \ref{fig:blockingarea2}.  \\ \hline
$\beq$ & A parameter defined to be equal to $\frac{\beta}{2\sqrt{\lbs\pi}}. $ \\ \hline
$\sbangle$ & The blocking cone created by the body of the user. \\ \hline
		\end{tabulary}
\end{table}
}

\begin{document}
\title{Macro diversity in Cellular Networks with Random Blockages}  \author{Abhishek K. Gupta, Jeffrey G. Andrews, and Robert W. Heath, Jr. \thanks{A. K. Gupta (g.kr.abhishek@utexas.edu), J. G. Andrews (jandrews@ece.utexas.edu) and R. W. Heath Jr. (rheath@utexas.edu) are with Wireless Networking and Communication  Group, The University of Texas at Austin, Austin,
TX 78712 USA.} \thanks{ This work is supported in part by the National Science Foundation under Grant 1514275 and AT\&T Laboratories.}}
\maketitle

\begin{abstract} 
Blocking objects (blockages) between a transmitter and receiver cause wireless communication links to transition from line-of-sight (LOS) to non-line-of-sight (NLOS) propagation, which can greatly reduce the received power, particularly at higher frequencies such as millimeter wave (mmWave).   We consider a cellular network in which a mobile user attempts to connect to two or more 
base stations (BSs) simultaneously,  to increase the probability of at least one LOS link, which is a form of macrodiversity.  We develop a framework for determining the LOS probability as a function of the number of BSs, when taking into account the correlation between blockages: for example, a single blockage close to the device -- including the user's own body -- could block multiple BSs.   We consider the impact of the size of blocking objects on the system reliability probability and show that macrodiversity gains are higher when the blocking objects are small.  We also show that  the BS density must scale as the square of the blockage density to maintain a given level of reliability.
\end{abstract}


\section{Introduction} \label{sec:Intro}

Blocking objects -- blockages -- in the form of buildings, foliage, and people, can severely impact the performance of cellular networks by reducing the signal strength and thus SNR.  Blocking's effect is more severe  at higher frequencies including mmWave, due to higher penetration losses and reduced diffraction \cite{Bai2014,Rappaport2013}. Therefore, LOS connections are highly desirable particularly for mmWave.   In addition, a user can block the otherwise LOS  signals due to its own body \cite{Baiselfblocking,Kiran2016}, hurting the overall reliability of the communication links. To overcome blockage effects, macrodiversity can be leveraged whereby a user is connected to multiple BSs simultaneously, which clearly increases the chance for a LOS connection \cite{Ghosh2014,Zhang2012}. 
The presence of large blockages results in correlation in the probability of NLOS propagation among BSs in the same general direction from the user, since the same object could block several LOS paths. The objective of this paper is to study the impact of blockages on the gains of macrodiversity including a realistic correlation model.
\subsection{Related Work}

There are some  approaches to  handle the blockage problem and increase link reliability in cellular systems,  in particular for mmWave frequencies. For example, using reflections from walls and other surfaces to steer around obstacles \cite{Genc2010} or  switching the beam from a LOS link to a NLOS link \cite{An2009} can reduce impact of blockages. 
However, this reduces received power vs a LOS link. 
 Another approach is to maintain link connectivity by use of relays and routing algorithms \cite{Singh2009,Niu2015}. However, this leads to other issues such as high latency and non-tractability, due to its complex algorithms and scheduling schemes. The third approach is to use macro diversity with multiple BSs \cite{Zhu2009,Zhang2012}.  Macro diversity over shadowing fading was studied in \cite{Shankar2008,Mukherjee2003}.  
In \cite{LeeMor2015}, the performance of coordinated beamforming with dynamic BS clusters was studied. The work \cite{HwangChae2013}  studied the fundamental limits of cooperation  for multicell cooperative networks with multiple receive antennas. In \cite{Zhang2012}, the authors proposed a multi-BS architecture for 60 GHz WLAN in which a MAC layer access controller device  is employed to enable each station to associate and cooperate with multiple BSs. In the proposed architecture, when one of wireless links is blocked, another BS can be selected to complete the remaining transmissions.

  In the past,  multiple approaches have been proposed to model and analyze blockages.
  Simulation based  approaches to model blockages  by using ray tracing \cite{Rizk1997} in a deterministic environment are numerically complex and not tractable. For tractability, blocking   
     is often included in the shadowing model as an additional loss. This approach, however, is over-simplistic, for example, it does not include the impact of the length of a link over its blocking probability. Therefore, this approach may not be not suitable to analyze scenarios where blockages play a significant role in determining its performance which is the case with communication at higher frequencies including mmWave.    In \cite{BaiLineBlock2012}, a tractable approach using random Boolean model with linear segments \cite{Cowan1989,ParkerCowan1976,PenStoyan89} was proposed to model the random blockages in a cellular system. This model was extended in \cite{BaiVaze2014} to include rectangular blockages and in \cite{Kiran2016} to include circular blockages. In \cite{Bai2014}, this blockage model was incorporated in the analysis of cellular systems to study the impact of blockages on the system performance. It was shown in \cite{Bai2014} that the link reliability and coverage probability depend on the blockage process as a function of the product of blockage density and average blockage length.

  Recent analytical work \cite{Bai2014,Kiran2016,BaiVaze2014} to analyze the impact of blockages in cellular systems assumes a single active link per user and does not include macro diversity.  In \cite{Choi2014}, a stochastic geometry framework was used to derive macro diversity gain for mmWave system in presence of random blockages. It assumes, however,  independence among blocking events of the different links. When simultaneous multiple links are considered, the larger blockages may decrease the diversity gains  due to induced correlation in blocking of these links and the system performance may no longer remain just a function of the product of the blockage density and average blockage length. In \cite{Samuylov2016}, time correlation of blocking events caused by user mobility was studied for mmWave networks. Although the spatial and time correlation analysis are similar to each other (in the sense that the correlation among  blocking of the  link to the same user at two time stamps is similar to the correlation among blocking of the links to different BSs), there are some differences in the analysis, insights and the interpretation. Also, in \cite{Samuylov2016}, the transmitter to receiver distance is fixed, which may not be the case with random deployment of BSs and users. Characterizing the spatial correlation among blockages and studying its impact in a system with macro diversity is the main focus of this work.


\subsection{Contributions}
In this paper, we  evaluate the benefits of macro diversity for a mmWave cellular system in the presence of random blockages. The contributions of the paper are summarized as follows.

\textbf{Analytical framework for dependent blocking.} We present a framework to analyze the correlation of blockage events occurring in links from the user to multiple BSs  in a cellular system with random blockages. For  blockage processes  with linear blockages, we compute the joint probability of these  links being non-blocked. We consider a special case of blockages where blockages have uniformly distributed lengths and orientations and show that unlike the independent case, this probability depends on both the blockage density and maximum blockage length, not just the product of the two. We  show that increasing the maximum blockage length while keeping the product  constant increases the correlation among  blockage events occurring in  multiple links.

\textbf{Gains from macro diversity.} We use the proposed framework to evaluate  gains in reliability obtained by the use of macro diversity.  We consider a system where each user is connected to multiple BSs simultaneously. For this system, we compute the average probability of having at least one LOS BS out of all connected links (termed the reliability).  We term the gain in reliability achieved by the use of multiple BSs connections {\em macro diversity gain}. We show that the required BS density to achieve a certain level of reliability can be decreased significantly by maintaining multiple BS links simultaneously. The correlation in blockage events decreases the macro diversity gain in comparison to the case where blocking is independent among links. We also show that to maintain same level of reliability, the BS density must scale as square of the blockage density.

\textbf{Analyzing diversity gains in the presence of self-blocking.}  If the person using a mobile phone comes in between the serving BS and the mobile, its body can block signals from its own serving BSs which is known as self-blocking. This  can be modeled by a cone at the user which blocks all BSs lying that cone. We assume that these multiple BSs are selected in a way to avoid self-blocking of all BSs at any time and  derive the reliability for this case using stochastic geometry tools.

 The rest of the paper is organized as follows. Section \ref{Sec:SysMod} explains the blockage and connectivity model. Section III considers a system with second order diversity and derive reliability for this system.  Section IV extends the analysis to the general case of the $n\ths$ diversity order.  Section V presents numerical results and explains the main insights of the paper. We  conclude in Section \ref{Sec:Conclusions}.

\section{System Model}\label{Sec:SysMod} 
In this section, we describe  three distinct aspects of the system model.

\textbf{Network model.}  
We consider a   cellular network consisting of BSs whose locations are modeled as a homogeneous Poisson point process (PPP)  with density $\lbs$ and users with locations modeled as a stationary Point process (PP). We consider a typical user at the origin $\mathbf{O}$.   Let $\PBS=\{\mathbf{x}_i, i\in \mathbf{N}\} $ denote the BS PPP where locations $\mathbf{x}_i$ are ordered according to their distances $R_i$ from the typical user. 

\begin{figure}[ht!]
\begin{center}
\subfigure[center][]{
\includegraphics[width=2.8in]{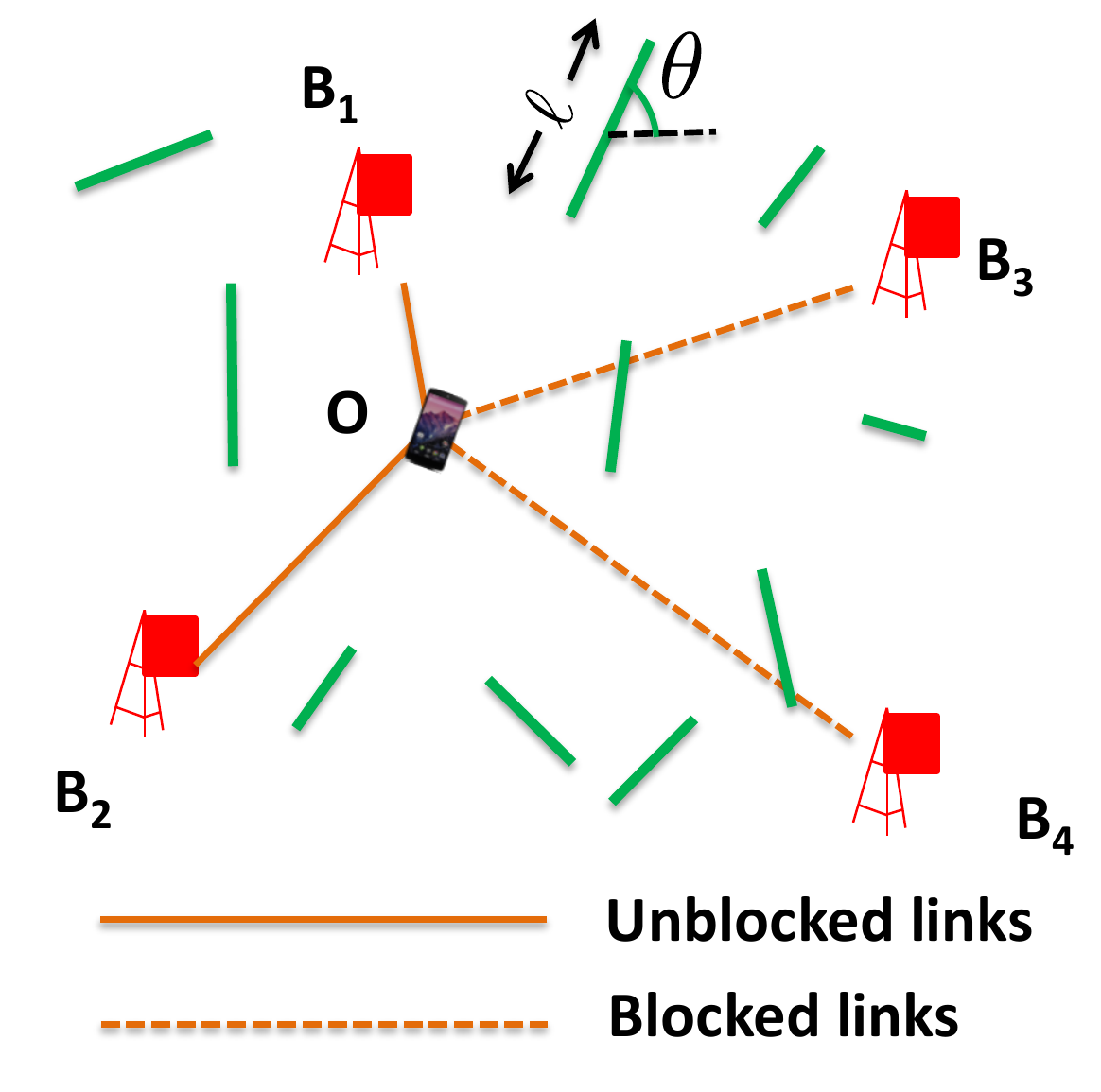}
\label{fig:sysmod}
}
\subfigure[center][]{\includegraphics[width=2.4in,trim=0 -80 0 0, clip=true ]{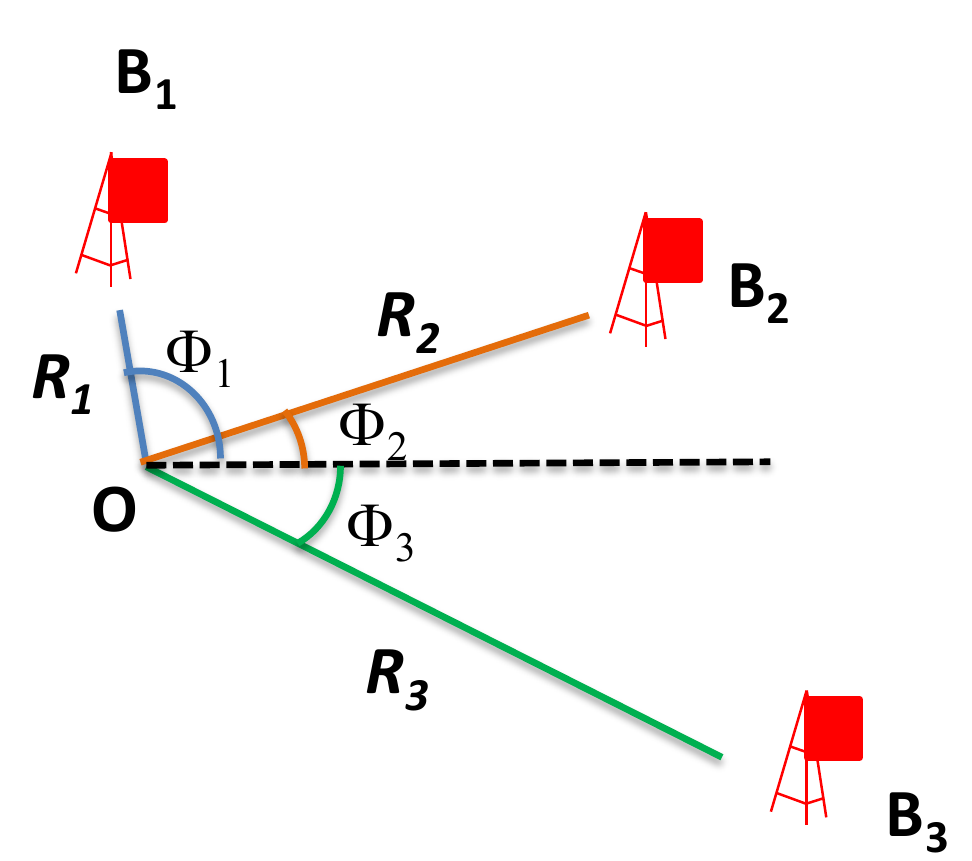}
\label{fig:sysmod3}}
\end{center}
\caption{System model: (a) An illustration showing a user at the origin in presence of blockages. Each blockage is modeled as  linear segment of length $\ell_k$ and orientation $\theta_k$. A link to any BS is said to be blocked or unreachable if a blockage falls on the link. (b) A cellular system with third order macro diversity ($n=3$). The typical user at $\mathbf{O}$ has three active BSs it can associate with.} 
\end{figure}

\textbf{Modeling random blockages.}  
 To model the blockages present in the channel, we consider the line Boolean model $\Psi$ similar to \cite{BaiLineBlock2012}. In this model, we assume that all the blockage elements are in the form of lines. In a real scenario, the blockages are polygon shaped. Since we are interested in their one dimensional intersections with the links, assuming their shapes as lines is a reasonable approximation. We also validate this model with real building data in Section \ref{sec:results}. The centers of these lines are modeled as a homogeneous PPP $\PBuild$ of density $\lbuild$. The lengths $\ell_k$ of the blockage lines are independent identically distributed (iid) random variables with distribution $F_L(\cdot)$. The orientations $\theta_k$ of the blockage lines are assumed to be iid random variables with distribution $F_\Theta(\cdot)$.  Let us define the average LOS radius  of a blockage process as $1/\beta$ where $\beta=\mu \frac{2}{\pi} \mathbb{E}[\ell]$ \cite{BaiVaze2014}. The analysis performed in this paper can  be in principle extended to Boolean models with other shapes such as rectangles or circles \cite{BaiVaze2014,Kiran2015Asilomar}, but
it is left for future work. In this paper, we will also consider a special case of the above blockage process  where $\ell$ and  $\theta$ are uniformly  distributed {\em i.e. }  $\ell\sim U(0,\Lmax), \theta\sim U(0,\pi)$   which is termed a {\em uniform blockage process}.

\textbf{Connectivity model.} 
We assume that all the users are simultaneously connected to the $n$ closest BSs  (see Fig. \ref{fig:sysmod}) where $n$ is  the {\em macro diversity order}. We assume that a user will be able to quickly establish communication links with any of the $n$ BSs connected to it, using a well designed initial access process. Recall that the link to a BS can be LOS if there are no  blockages intersecting the link between the BS and the user, otherwise the link is said to be in NLOS. Let  $\mathcal{A}_i$ denote the event that the $i\ths$ BS is LOS. At any point in time, the user will establish a communication link with the closest LOS BS, termed {\em the associated BS} out of these $n$ connected BSs. If all $n$ of the closest BSs are blocked, we say that the user is fully blocked.

We define the reliability $\pR$  as the probability that at least one connected BS out of the $n$ connected BSs is LOS to a typical user and is given as
\begin{align}
\pR&=\prob{\bigcup_{i=1}^n \mathcal{A}_i} \label{eq:pRdef}.
\end{align}
The reliability is  useful from a system point of view. For example, a cellular operator may be interested in the question that if each user can use $n\ths$ order macro diversity, what the required BS density should be to achieve certain level of reliability. 

\insertnotationtable


\section{Reliability Analysis for Second Order macro diversity}
We will now compute the reliability for the system model described in the last section. To simplify, we will first consider a system of second order macro diversity ($n=2$) and we will then extend the analysis to the general $n$ case.  In the $n=2$ case, the typical user is connected with two BSs, $\point{B}_1$ and $\point{B}_2$ with link lengths equal to $R_1$ and $R_2$. Let us denote the angle between the two links $\point{Z}_1$ and $\point{Z}_2$ as $\Phi$. Without loss of generality, we assume that $\mathbf{x}_2$ is at $x$ axis and $\Phi\ge0$. The joint distribution of $R_1$ and $R_2$ is given as
\begin{align}
f(r_1,r_2)&=(2\pi\lambda)^2r_1r_2\exp(-\lambda\pi r_2^2), \text{ if }r_1\le r_2 \label{eq:jointn2r1r2}
\end{align}
and $\Phi\sim \text{Uniform}(0,\pi)$.

Now, let us consider a derived Boolean model $\Psi(\ell,\theta)$  consisting of the blockages in $\Psi$ with lengths between $\ell$ and $\ell+\dd\ell$ and orientations between $\theta$ and $\theta+\dd\theta$. Note that the process containing the centers of these blockages can be obtained by thinning the original PP $\PBuild$. Therefore, centers of $\Psi(\ell,\theta)$  form a PPP with intensity $\lbuild F_L(\dd\ell)F_\Theta(\dd\theta)$. Given the two links $\point{Z_1}$ and $\point{Z_2}$, let $\event{A}_1$ and $\event{A}_2$ respectively denote the events that these links  are unblocked.  The probability that at least one of the links is not blocked is given as
\begin{align}
\prob{\event{A}_1\cup\event{A}_2}=\prob{\event{A}_1}+\prob{\event{A}_2}-\prob{\event{A}_1\cap\event{A}_2}.
\label{eq:pacupb}
\end{align}

\begin{figure}[ht!]
\begin{center}
\includegraphics[width=5.9in,trim=0 00 0 0,clip]{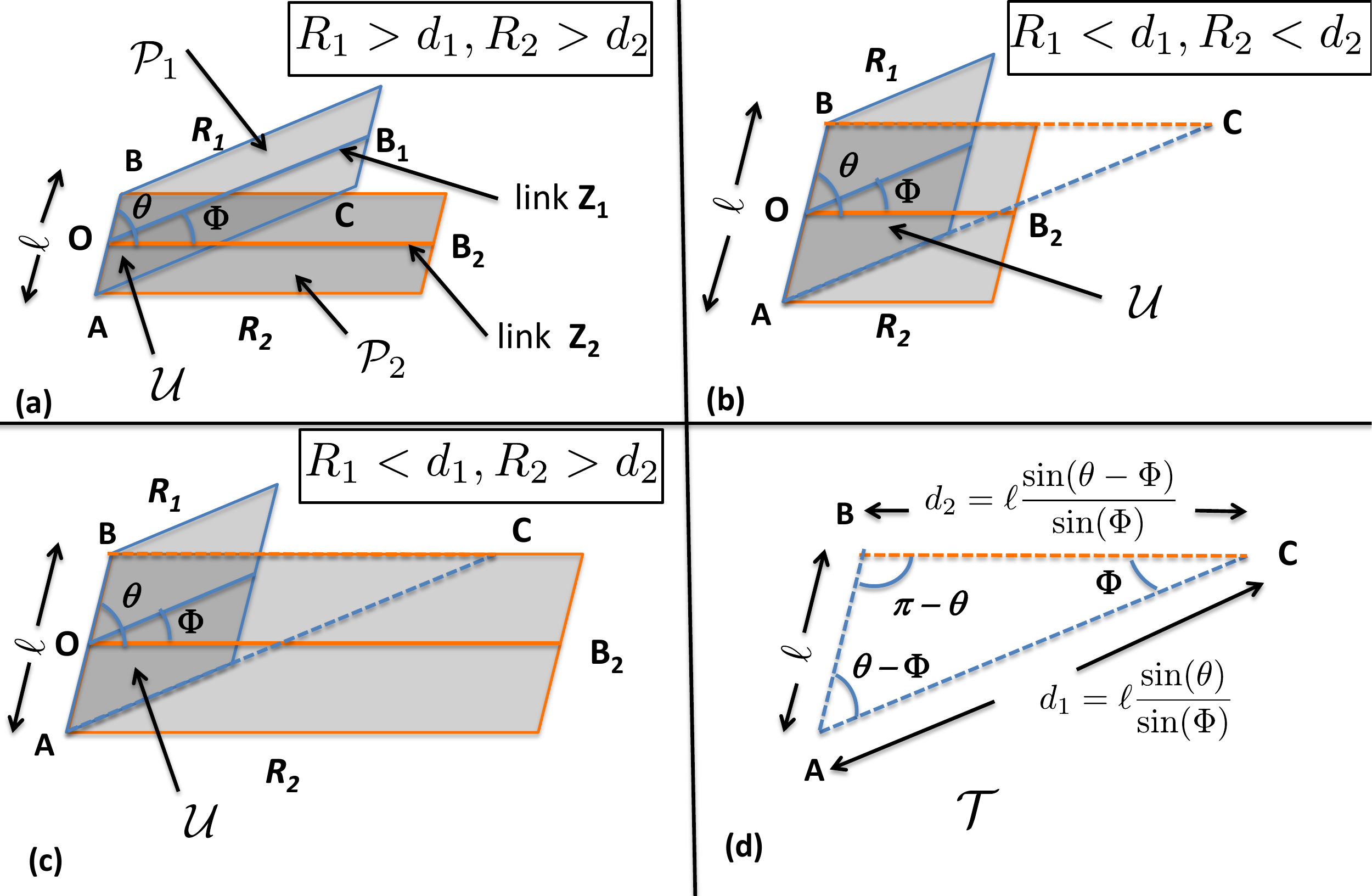}
\caption{Illustration showing the joint area of the two parallelograms $\mathcal{P}_1$ and $\mathcal{P}_2$. The parallelogram $\mathcal{P}_1$ (or $\mathcal{P}_2$) represents the region where centers of any blockage must not lie for the links $\point{Z}_1$ (or $\point{Z}_2$) to be LOS. Subfigures (a), (b) and (c) show  shapes for the region $\mathcal{U}$ common in both parallelograms dependent on the values of $\ell,\theta,R_1,R_2$ and $\Phi$. The conditions are mentioned in the subfigures where $d_1$ and $d_2$ are  lengths of AC and BC. Fig. (d) The triangle $\mathcal{T}$ circumscribes the common region $\mathcal{U}$. }
\label{fig:blockingarea2}
\end{center}
\end{figure}

The event  that the link $\point{Z}_1$ is not blocked by a blockage in $\Psi(\ell,\theta)$ is equivalent to the event that the centers of all blockages in this collection  $\Psi(\ell,\theta)$ lie outside the parallelogram $\mathcal{P}_1$ shown in Fig. \ref{fig:blockingarea2}(a). The area of $\mathcal{P}_1$ is given as
\begin{align*}
A_1(R_1,\ell,\theta,\Phi)&=\ell R_1 \sin(|\theta-\Phi|).
\end{align*}
 Hence, the probability of the event that the link $\point{Z}_1$ is not blocked by an blockage in $\Psi(\ell,\theta)$ is given by void probability of a PPP and is equal to $\exp\left(-\lbuild F_L(\dd\ell)F_\Theta(\dd\theta) A_1(R_1,\phi,\ell,\theta)\right)$. 
Therefore, the probability of the event $\event{A}_1$ that the link $\point{Z}_1$ is not blocked is
\begin{align}
\prob{\event{A}_1}&=\prod_{\ell,\theta}\exp(-\lbuild F_L(\dd\ell)F_\Theta(\dd\theta) A_1(R_1,\phi,\ell,\theta))\nonumber\\
&=\exp\left(-\int_0^\infty \int_0^\pi\lbuild  A_1(R_1,\phi,\ell,\theta)F_L(\dd\ell)F_\Theta(\dd\theta)\right)\label{eq:pa1s1}.
\end{align}
Using the value of area $A_1$ in \eqref{eq:pa1s1}, we get
\begin{align}
\prob{\event{A}_1}&=\exp\left(-\int_0^\infty \int_0^\pi\lbuild \ell R_1 \sin(|\theta-\Phi|)  F_L(\dd\ell)F_\Theta(\dd\theta)\right)\nonumber\\
&=\exp\left(-\lbuild\expect{\ell}R_1\frac2\pi\right)=\exp(-\beta R_1)
\label{eq:prexp1}.
\end{align}
Similarly the probability of the event $\event{A}_2$ that the link $\point{Z}_2$ is not blocked  is
\begin{align}
\prob{\event{A}_2}
&=\exp\left(-\int_0^\infty \int_0^\pi\lbuild  A_2(R_2,\phi,\ell,\theta)F_L(\dd\ell)F_\Theta(\dd\theta)\right)=\exp(-\beta R_2)
\label{eq:prexpB}.
\end{align}

Now we will compute the joint probability $\prob{\event{A}_1\cap\event{A}_2}$. Similar to the previous case,  the event that both links are not blocked by any blockage in $\Psi(\ell,\theta)$   is equivalent to the event that centers of all blockages in the collection  $\Psi(\ell,\theta)$ lie outside the shaded region  (which is the union of two parallelograms $\mathcal{P}_1$ and $\mathcal{P}_2$)  shown in Fig. \ref{fig:blockingarea2}. Let $A(R_1,R_2,\phi,\ell,\theta)$ be the area of shaded region. The shape of the intersection (denoted by $\mathcal{U}$)  of the two parallelograms can be triangular or trapezoidal, dependent on the values of $R_1,R_2,\Phi,\theta,\ell$ (See Fig. \ref{fig:blockingarea2}(a-c)). Let triangle $\mathcal{T}$ denote the triangle  ($\Delta$ABC) circumscribing $\mathcal{U}$. The area of this triangle is 
\begin{align*}
T=\frac{\ell^2\sin(\theta)\sin(|\theta-\Phi|)}{2\sin(\Phi)}.
\end{align*}
 The trapezoidal shape occurs only when $R_1<\ell_1$ or $R_2<\ell_2$. The area of $\mathcal{T}\setminus\mathcal{U}$ is given by $\left(1-\min\left(\frac{R_2}{\ell_1},\frac{R_1}{\ell_2}\right)\right)^2T$. Therefore,  the area of  $\mathcal{P}_1\cup\mathcal{P}_2$ is
\begin{align}
&A(R_1,R_2,\Phi,\ell,\theta)=\ell R_1 \sin(|\theta-\Phi|)+\ell R_2 \sin(\theta)\nonumber\\
&\ \ \ -1(\theta>\phi)\frac{\ell^2\sin(\theta)\sin(\theta-\Phi)}{2\sin(\Phi)}\left[1-\left(1-\min\left(1,\frac{R_1\sin(\Phi)}{\ell\sin(\theta)}
\frac{R_2\sin(\Phi)}{\ell\sin(\theta-\Phi)}
\right)\right)^2\right]. \label{eq:areaexp}
\end{align}
 Hence, the probability that both of the links are not blocked  by $\Psi(\ell,\theta)$ is 
$\exp\left(-\lbuild F_L(\dd\ell)F_\Theta(\dd\theta)\right.$   $\left. A(R_1,R_2,\phi,\ell,\theta)\right)$.
 Therefore, the probability that both links are not blocked  is 
\begin{align}
\prob{\event{A}_1\cap \event{A}_2}
&=\prod_{\ell,\theta}\exp(-\lbuild F_L(\dd\ell)F_\Theta(\dd\theta) A(R_1,R_2,\phi,\ell,\theta))\nonumber\\
&=\exp\left(-\int_0^\infty \int_0^\pi\lbuild  A(R_1,R_2,\phi,\ell,\theta)F_L(\dd\ell)F_\Theta(\dd\theta)\right)
\label{eq:prexp}
\end{align}
Let us define $\s(R_1,R_2,\Phi)$ as the mean shaded area averaged over blockage size and orientation distribution:
\begin{align}
\s(R_1,R_2,\Phi)&=\int_0^\infty \int_0^\pi  A(R_1,R_2,\phi,\ell,\theta)F_L(\dd\ell)F_\Theta(\dd\theta).
\label{eq:prexps}
\end{align}
Then, $\prob{\event{A}_1\cap \event{A}_2}$ is given by
\begin{align}
\prob{\event{A}\cap \event{A}_2}=p(R_1,R_2,\Phi)
&=\exp\left(-\lbuild \s(R_1,R_2,\Phi) \right)
\label{eq:prexp23}.
\end{align}
Therefore, using \eqref{eq:pacupb}, the probability that at least one of the links is not blocked is equal to
\begin{align}
\prob{\event{A}_1\cup\event{A}_2}=\exp(-\beta R_1)+\exp(-\beta R_2)-\exp\left(-\lbuild \s(R_1,R_2,\Phi) \right).
\label{eq:pacupb2}
\end{align}
Now, the reliability can be computed as
\begin{align}
&\pR=\expects{R_1,R_2}{\prob{\event{A}_1\cup\event{A}_2}}\nonumber\\
&=\frac1\pi\int_0^\infty\int_0^\infty\int_0^\pi\left(\exp(-\beta r_1)+\exp(-\beta r_2)-\exp\left(-\lbuild \s(r_1,r_2,\phi) \right)\right) f_{R_1,R_2}(r_1,r_2)\dd\phi\dd r_1\dd r_2\nonumber\\
&=\frac{(2\pi\lambda)^2}\pi\int_0^\infty\int_0^{r_2}\int_0^\pi\left(\exp(-\beta r_1)+\exp(-\beta r_2)\right.\nonumber\\
&\hspace{2in}\left.-\exp\left(-\lbuild \s(r_1,r_2,\phi) \right)\right)r_1r_2\exp(-\lambda\pi r_2^2) \dd\phi\dd r_1\dd r_2.
\label{eq:pacupb2}
\end{align}
Using the transformations $x_1=\beta r_1,x_2=\beta r_2$,
\begin{align}
&\pR=\frac1{4\beq^4\pi}\int_0^\infty\int_0^{x_2}\int_0^\pi\left(\expU{-x_1}+\expU{-x_2}-\expU{-\lbuild s\left(\frac{x_1}{\beta},\frac{x_2}{\beta},\phi\right)}\right)x_1x_2\expU{-\frac{x_2^2}{4\beq^2}} \dd\phi\dd x_1\dd x_2
\label{eq:pacupb3}
\end{align}
where  
\begin{align}
\beq=\frac{\beta}{2\sqrt{\pi\lambda}}.
\end{align} 
Here, $\beq$ equals half the ratio of average cell radius ($ 1/\sqrt{\pi \lbs} $) and average LOS radius ($1/\beta$) and therefore, represents the relative blockage size with respect to the BS deployment. Solving the first two integrals in \eqref{eq:pacupb3}, we get the final expression for the reliability given in the following Theorem. 
\begin{theorem}\label{thm:n2DEP}
The reliability  in a cellular network with second order  macro diversity is 
\begin{align}
\pR&=2+\beq^2-\beq(5+2\beq^2)\Q(\beq)-\nonumber\\
&\frac{1}{4\beq^4\pi}\int_{0}^{\pi}\int_{0}^{\infty} x_2\exp\left(-\frac{x_2^2}{4\beq^2}\right) \int_{0}^{x_2} \exp\left(-\lbuild \s\left(\frac{x_1}{\beta},\frac{x_2}{\beta},\phi\right)\right) x_1 \dd x_1 \dd x_2 \dd\phi
\end{align}
where $\s(r_1,r_2,\phi) $ is given in \eqref{eq:prexps}, and 
\begin{align}\label{eq:Qfunc}
\Q(x)=\frac{\sqrt{\pi}}2\mathrm{erfcx}(x)
=\sqrt{\pi}\exp{(x^2)}\mathrm{Q}\left(\sqrt{2}x\right).
\end{align}
\end{theorem}
Before going further, we will give the following Lemma regarding the monotonicity of reliability with respect to blockage length's distribution. 

\begin{lemma}\label{lemma:monotone}
For all the  blockage processes with the same parameter $\beta$ and  scaled  distribution $F'_c(\dd \ell)=F_L(\dd \ell/c)$ of length, the quantity $\lbuild \s(R_1,R_2,\Phi)$, as defined in \eqref{eq:prexps}, monotonically decreases with increasing $c$ and so does the reliability.
\end{lemma}

\begin{IEEEproof}
See Appendix \ref{appen:prooflemmamonotone}.
\end{IEEEproof}

Note that for  the uniform blockage process with blockage length $\ell\sim U(0,\Lmax)$, scaling the distribution by $c$, is equivalent to increasing the maximum length $\Lmax$ by $c$.
Therefore, one direct result of the  Lemma \ref{lemma:monotone} for  the uniform blockage process  is that the reliability decreases with increasing $\Lmax$. This result is intuitive as fewer but more  bulky blockages  make the blocking probability of two links more correlated while small but more blockages result in independence between  blocking of any two links.

The function $\s(\cdot,\cdot,\cdot)$ in Lemma \ref{lemma:monotone} is dependent on the distribution of $\ell$ and $\theta$. In general, this function is difficult to compute. In the next subsections, we will consider a few special cases to simplify  the  function $\s(\cdot,\cdot,\cdot)$ to get closed form expressions for the reliability.



\subsection{Reliability for Independent Blocking}
In this subsection, we consider the independent blocking scenario where both links are blocked or not independently.  Then, $\prob{\event{A}_1\cap\event{A}_2}$ is equal to
\begin{align}
\mathbb{P}_{\mathrm{IND}}\left[\event{A}_1\cap\event{A}_2\right]
=\prob{\event{A}_1}\prob{\event{A}_2}=\exp(-\beta R_1-\beta R_2).
\end{align}

Note that the area of $\mathcal{P}_1\cup\mathcal{P}_2$ is greater than the sum of the areas of the two parallelograms  {\em i.e.}
\begin{align}
A(R_1,R_2,\Phi,\ell,\theta)&\le\ell R_1 \sin(|\theta-\Phi|)+\ell R_2 \sin(\theta).\nonumber
\end{align}
Averaging with respect to  $\ell$ and $\theta$, we can upper bound mean shaded area as
\begin{align}
\s(R_1,R_2,\Phi)&\le \lbuild \expect{\ell}\frac{2}\pi (R_1+R_2)\nonumber\\
\prob{\event{A}_1\cap\event{A}_2}&=\exp(-\lbuild \s(R_1,R_2,\Phi))\ge\exp(-\beta R_1-\beta R_2)=\mathbb{P}_{\mathrm{IND}}\left[\event{A}_1\cap\event{A}_2\right].
\end{align}
Therefore, the independent blocking case upper bounds the reliability in the dependent blocking scenario. Hence, we denote it by the notation $\pRIND$. We now provide  the exact expression of  the reliability for the independent blocking case  in the following Theorem.
\begin{theorem} \label{thm:n2IND}
The reliability  in a cellular network with second order  diversity and independent blocking is 
\begin{align}
\pRIND&=\frac{1}{\beq}\left[\beq^3-\Q(\beq)(2\beq^4+5\beq^2-1)+\Q(2\beq)(8\beq^2-1)\right]
\end{align}
where $\Q(\cdot)$ is given in \eqref{eq:Qfunc}.
\end{theorem}
\begin{IEEEproof}
See Appendix \ref{appen:thmn2IND}.
\end{IEEEproof}
The above Theorem directly gives the following Corollary.

\begin{corollary}
Given a certain target value of $\pRIND$, the required BS density  with independent blocking is given by $\lambda=\frac{\beta^2}{4\pi \beq_s^2}$ where $\beq_s$ is the solution of the following equation
\begin{align}
\beq^3-\Q(\beq)(2\beq^4+5\beq^2-1)+\Q(2\beq)(8\beq^2-1)-\beq\pRIND=0.\label{eq:reversen2}
\end{align}
\end{corollary}
\noindent Given $\pRIND$, \eqref{eq:reversen2} can be solved for $\beq$ using a numerical method.

\subsection{Reliability for  Uniform Blockage Process} 
In this subsection, we will consider a uniform blockage process  and provide  bounds for the reliability.  In a uniform blockage process, the length and orientation of blockages are uniformly distributed. For this case, the average LOS radius is given as $\beta=\lbuild \Lmax/\pi$ where $\Lmax$ is the maximum length of blockages.

\subsubsection{Lower Bound I} Note that the area $A$ is the sum of $A_1$ (the area of $\mathcal{P}_1$) and $A_2$ (the area of $\mathcal{P}_2$) minus the area of the common region $\mathcal{U}$ (in shape of either a trapezoid or a triangle). We can lower bound this area $A$ by replacing the area of $\mathcal{U}$ by the area of its circumscribing triangle $\mathcal{T}$. Note that for certain values of $\theta$ and $\ell$, the area of $\mathcal{T}$  may become greater than the area of parallelogram $\mathcal{P}_1$. In this case, we can lower bound the area $A$ by just the area of parallelogram $\mathcal{P}_2$. Hence, we get the following lower bound for area $A$:
\begin{align}
A(R_1,R_2,\phi,\ell,\theta)&\ge
\begin{cases}
A_1+A_2 & \text{ for } \theta\le\phi \\
A_1+A_2-T & \text{for } \pi>\theta>\phi, \sin(\theta)<{2R_1\sin(\phi)}/{\ell}\\
A_2 & \text{for } \pi>\theta>\phi, 1\ge\sin(\theta)>{2R_1\sin(\phi)}/{\ell}.
\end{cases}\label{eq:Alowerbound}
\end{align}
\renewcommand{\a}{a}
Now, integrating \eqref{eq:Alowerbound} with respect to distribution of $\theta$ and $\ell$ gives the lower bound for $\s(R_1,R_2,\Phi)$ which is denoted by $\LBI{\s}(R_1,R_2,\Phi)$ and given as
\begin{align*}
\s(R_1,R_2,\Phi)\ge \LBI{\s}(R_1,R_2,\Phi)=\frac{\Lmax}{\pi} \left(R_1+R_2- R_1 F\left(\frac{R_1}{\Lmax},\Phi \right)\right)
\end{align*}
 where $F(a,\Phi)=$
\begin{align*}
&
\begin{cases}0<\Phi\le\frac\pi2&
\begin{cases}
 2\a\le 1& \frac12+\frac{\a}3(2\a-3)\sin^2(\Phi)+\frac{1}3T_1-\frac23T_2+\frac43T_3-\frac23T_4+\frac1{12}T_6 \\
2\a>1,2\a\sin(\Phi)\le 1& \frac12-\a(2\a+1)\sin^2(\Phi)+\frac13T_1-\frac23T_2+\frac23T_3+\frac{1}{12}T_6+T_7\\
2\a\sin(\Phi)\ge 1&
\frac{1}{12\a}(1+(\pi-\Phi)\cot(\Phi)) 
\end{cases} \\
\frac\pi2<\Phi<\pi&
\begin{cases}
2\a\le 1 &\hspace{1in}
\frac12  +\frac{\a}{3}(2\a-3)\sin^2(\Phi)+\frac13 T_1 -\frac23T_2  +\frac23T_5 +\frac1{12}T_6  \\
2\a>1& \hspace{1in}
\frac{1}{12\a}(1+(\pi-\Phi)\cot(\Phi)) 
\end{cases} 
\end{cases} 
\end{align*}
%
%
%
%
%
%
%
with
\begin{align*}
T_1&=\cos(\Phi)\sqrt{1-{4\a^2\sin^2(\Phi)}}&
T_2&=\a^2\sin^2(\Phi)\cos(\Phi)\log\left(1+\sqrt{1-4\a^2\sin^2(\Phi)}\right) \\
T_3&=\a^2\sin^2(\Phi)\cos(\Phi)\log(2\a\sin(\Phi))&
T_4&=\a^2\sin^2(\Phi)\cos(\Phi)\log(2\a(1+\cos(\Phi)))\\
T_5&=\a^2\sin^2(\Phi)\cos(\Phi)\log(2\a(1-\cos(\Phi)))&
T_6&=\frac{1}{\a}\cot(\Phi)\sin^{-1}\left(2\a\sin(\Phi)\right)\\
T_7&=\frac{(\pi-2\Phi)}{3}a^2\sin^2(\Phi)\cos(\Phi).&
\end{align*}
Note that  $F(a,0)=1$ and $F(a,\pi)=0$.

It can be seen that the lower bound on the mean area  $\LBI{\s }(R_1,R_2,\Phi)$ is dependent on both the blocking parameter $\beta$ and the maximum blockage length $\Lmax$.   The monotonicity of $A$ with respect to $\Lmax$ (as shown in Lemma \ref{lemma:monotone}) implies that for a constant $\beta$, as $\Lmax$ increases (which means fewer  but larger blockages), $F(\a,\Phi)$ increases towards $\frac12(1+\cos(\Phi))$. For small $\Lmax$, (which means more  but smaller blockages), $F(\a,\Phi)$ decreases to $0$ which corresponds to the independent case. For intermediate values of $\Lmax$, $F(\a,\Phi)$ will range between $0$ and $\frac12(1+\cos(\Phi))$.

Now, the lower bound on reliability  can be obtained by using Theorem \ref{thm:n2DEP}:
\begin{align*}
\pR \ge &\  \LBI{\pR}=2+\beq^2-\beq(5+2\beq^2)\Q(\beq)\\
&-\frac{1}{2\beq^2\pi}\int_{0}^{\pi}\int_{0}^{\infty}  \exp(-x_1(1-F(x_1/(\beta \Lmax),\phi))) x_1S(x_1) dx_1  \dd\phi
\end{align*}
where $S(x)=\exp\left(\beq^2-\left(\frac{1}{2\beq}x+\beq\right)^2\right)\left[1-2\beq\Q\left(\frac{1}{2\beq}x+\beq\right)\right]$.

\subsubsection{Asymptotic Lower Bound}
We can  bound the Area $A$ as follows:
\begin{align}
A(R_1,R_2,\phi,\ell,\theta)&\ge
\begin{cases}
A_1+A_2 & \text{ for } \theta\le\phi \\
A_2 & \text{for } \pi>\theta>\phi
\end{cases}\label{eq:ALowerBound2}.
\end{align}
Now, integrating \eqref{eq:ALowerBound2} with respect to $\theta$ and $\ell$ gives the following lower bound (denoted by $\LBII{\s}$) for mean area $\s(R_1,R_2,\Phi)$:
\begin{align}
\s(R_1,R_2,\Phi)\ge \LBII{\s}(R_1,R_2,\Phi)=\frac{\Lmax}{\pi} \left(R_1+R_2- R_1 \frac{1}{2}\left(1+\cos(\Phi) \right)\right)\label{eq:sLowerBound2}.
\end{align}
As discussed in the previous subsection, for a given $\beta$, the lower bound becomes asymptotically tight as maximum blockage length $\Lmax\rightarrow\infty$.  Now, using the lower bound in \eqref{eq:sLowerBound2} and Theorem \ref{thm:n2DEP}, a lower bound (denoted by $\pRDEP$) on reliability probability can be obtained as follows (see Appendix \ref{appen:proofpRDEP2}):
\begin{align}
\pR&\ge \pRDEP= 1+\beq^2-\beq(5+2\beq^2)\Q(\beq)+\frac{4\beq}\pi   
\int_{0}^{\pi/2}
(2+\sin^2(\phi)) \Q((1+\sin^2(\phi))\beq))\dd\phi
\nonumber\\
&+\frac2\pi   
\frac{1}{\beq}\int_{0}^{\pi/2}
\left[\Q(\beq)
- \Q((1+\sin^2(\phi))\beq))\right.\nonumber\\&\left.
+2\beq^2\sin^2(\phi)\Q((1+\sin^2(\phi))\beq))
-\sin^2(\phi)\beq\right]\mathrm{cosec}^4(\phi)\dd\phi\label{eq:pRLowerBound2}.
\end{align}

The lower bound given in \eqref{eq:pRLowerBound2} can also be  approximated  using the linear approximation:  $\sin^2{\Phi/2}\approx\Phi/\pi$ for $0\le\Phi\le\pi$ (see Appendix \ref{appen:proofpRDEPlinearapprox})
  \begin{align*}
\pRDEP{\approx}
\frac1\beq\left[3\beq+\beq^3-\Q(\beq)(2\beq^4+7\beq^2+2)+2\Q(2\beq)\right].
\end{align*}

Now using the bounds \eqref{eq:pRLowerBound2} along with Theorem \ref{thm:n2IND} and Lemma \ref{lemma:monotone}, we give the following Theorem. 
\begin{theorem}\label{thm:n2ULBound}
The reliability of a cellular system with second order macro diversity in the presence of a uniform blockage process  of parameter $\Lmax$ with fixed $\beta$, is bounded as
\begin{align}
\pRDEP(\beta/\sqrt{\lbs})
 \le \pR(\lbs,\lbuild,\Lmax)\le \pRIND(\beta/\sqrt{\lbs})
\end{align}
where the lower bound is tight for $\Lmax\rightarrow \infty$ and the upper bound for $\Lmax\rightarrow0$.
\end{theorem}

It can be seen from Theorem \ref{thm:n2ULBound} that reliability is bounded above and lower by  expressions which are  functions of only $\beta/\sqrt{\lbs}$. It implies that BS density $\lbs$ needs to scale as  $\beta^2$ (and hence as square of blockage density ($\lbuild^2$) for fixed maximum blockage length) to maintain the same order of reliability. This trend is consistent with a system with no diversity \cite{BaiVaze2014}.

\begin{figure}[ht!]
\begin{center}
\includegraphics[width=3in]{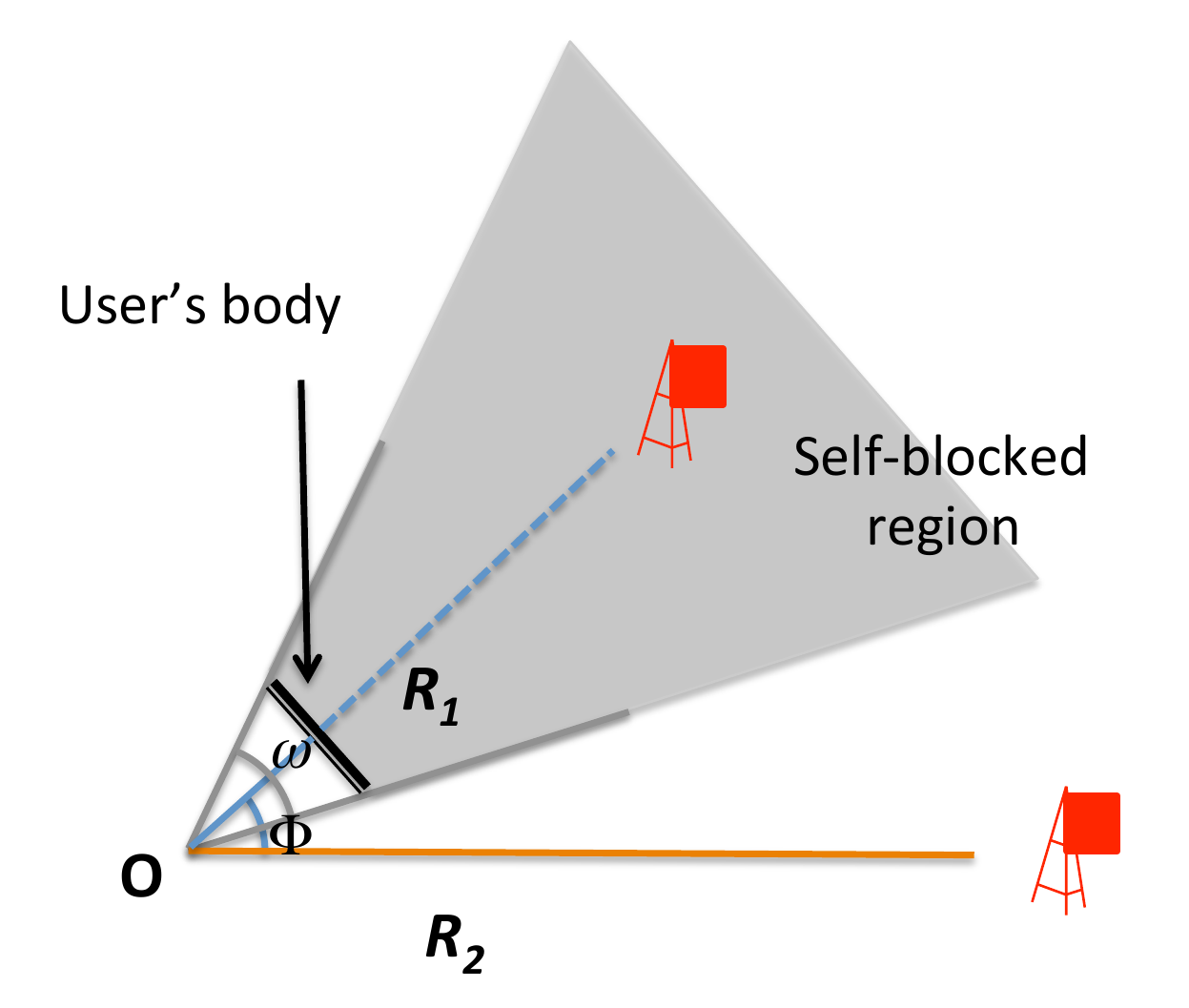}
\caption{Blocking cone created by the user's own body which can block its own serving BS.}
\label{fig:selfblocking}
\end{center}
\end{figure}

\subsection{Reliability Analysis Under Self-blocking}

In this subsection, we include  self-blocking in the analysis where a user can self-block its own serving BS. Self-blocking in a cellular network can be modeled using a blocking cone with angle $\sbangle$  in the body's direction which blocks all the BSs behind the body (see Fig. \ref{fig:selfblocking}) \cite{Kiran2016,Baiselfblocking}. The angle $\sbangle$ depends on the width of the user's body  and its distance from the mobile. In this case, we assume that when we select the two BSs for a user, they will be chosen in a way such that always, at least one of the two BSs is not blocked by its own body. This means that the angle between the two BSs must be more than $\sbangle$. In other words, given the closest BS at distance $R_1$, the second BS should be chosen such that the angle $\Phi$ between the two BSs satisfies $\pi\ge\Phi>\sbangle$ or $\pi\le\Phi<-\sbangle$. Let us denote the distance to the second BS is $D_2$. Note that $D_2$ is different than $R_2$ due to the constraint of this BS to be outside the self-blocking cone. We assume that the users' orientation is uniformly distributed between 0 and $2\pi$. Therefore, given the two selected control links, the marginal probability that each link is not self-blocked is given by $c=1-\frac{\sbangle}{\pi}$ and the joint probability that both links are not self-blocked is given by $c_2=1-\frac{2\sbangle}{\pi}$.

The joint distribution of $R_1$, $D_2$ and $\Phi$ can be computed as
\begin{align*}
&f_{R_1,D_2,\Phi}(r_1,r_2,\phi)=\\
&2\pi\lambda^2 r_1r_2\exp(-\lambda\pi c r_2^2-\lambda\pi (1-c) r_1^2)\indside{r_1\le r_2}\indside{(\pi\ge\phi>\sbangle)\cup(-\pi\le\phi<-\sbangle)}.
\end{align*}
Integrating  with respect to $\Phi$ and $R_1$, we  get the marginal distribution of $D_2$
\begin{align}
f_{D_2}(r_2)&=\frac{2c\pi\lbs r_2}{1-c}\left[\exp\left(-\lbs\pi c r_2^2\right)-\exp\left(-\lbs\pi r_2^2\right)\right].
\end{align}



Now, let $\event{A}_1$  be the event that the link $\point{\Z}_1$ is not blocked (neither blocked by a blockage or self-blocked). Similarly let  $\event{A}_2$ be the event that the link $\point{\Z}_2$ is not blocked. Then, $\prob{\event{A}_1}$ and $\prob{\event{A}_2}$ is given as
\begin{align}
\prob{\event{A}_1}&=c\exp(-\beta R_1),& \prob{\event{A}_2}&=c\exp(-\beta R_2).
\end{align}
and the joint probability of both links being unblocked is given as 
\begin{align}
\prob{\event{A}_1\cap\event{A}_2}&=c_2\exp(-\lbuild \s(R_1,D_2,\Phi)).
\end{align}
The reliability is given as
\begin{align}
\pR&=\expects{R_1,R_2}{\prob{\event{A}_1\cup\event{A}_2}}\nonumber\\
=&2\pi\lbs^2\int_0^\infty\int_0^{r_2}\int_0^{2\pi}\left(c\exp(-\beta r_1)+c\exp(-\beta r_2)-c_2\exp\left(-\lbuild \s(r_1,r_2,\phi) \right)\right)\nonumber
\\
&\exp(-\lambda\pi c r_2^2-\lambda\pi (1-c) r_1^2) \indside{(\pi\ge\phi>\sbangle)\cup(-\pi\le\phi<-\sbangle)}
\dd\phi\dd r_1\dd r_2
.\end{align}
Using the transformations $x_1=\beta r_1$ and $x_2=\beta r_2$, and noting the symmetry of inner term with respect to $\phi$ around x axis, 
\begin{align}
\pR
=&\frac1{4\beq^2\pi}\int_0^\infty\int_0^{x_2}\int_\sbangle^{\pi}\left(c\exp(-x_1)+c\exp(-x_2)-c_2\exp\left(-\lbuild \s(x_1/\beta,x_2/\beta,\phi) \right)\right)\nonumber\\
&\exp(-c x_2^2/(4\beq^2)-(1-c) x_1^2/(4\beq^2)) 
\dd\phi\dd x_1\dd x_2\nonumber\\
=&c\left(2-\frac{2\beq}{1-c}\left({\Q(\beq)}-\frac{\Q(\beq/\sqrt{c})}{\sqrt{c}}\right)\right)\nonumber-\frac{c_2}{4\beq^2\pi}\int_0^\infty\int_0^{x_2}\int_\sbangle^{\pi}\exp\left(-\lbuild \s(x_1/\beta,x_2/\beta,\phi) \right)\\
&\hspace{1in}\times \exp(-c x_2^2/(4\beq^2)-(1-c) x_1^2/(4\beq^2)) 
\dd\phi\dd x_1\dd x_2.\label{eq:sbpRDEP}
\end{align}

For the independent case, the reliability can be obtained by replacing function $\s(r_1,r_1,\phi)$ in \eqref{eq:sbpRDEP} by $\beta r_1+\beta r_2$:
\begin{align}
\pRIND
=&c\left(2-\frac{2\beq}{1-c}\left({\Q(\beq)}-\frac{\Q(\beq/\sqrt{c})}{\sqrt{c}}\right)\right)\nonumber-\frac{c_2c}{4\beq^2}\int_0^\infty\int_0^{x_2}\exp\left(-x_1-x_2 \right)\\
&\hspace{1in}\times \exp(-c x_2^2/(4\beq^2)-(1-c) x_1^2/(4\beq^2)) 
\dd x_1\dd x_2\nonumber\\
=&c\left[2-\frac{2\beq}{1-c}\left({\Q(\beq)}-\frac{\Q(\beq/\sqrt{c})}{\sqrt{c}}\right)-\frac{c_2}{4\beq^2}\int_0^\infty\int_0^{x_2}\exp\left(-x_1-x_2 \right)\right.\nonumber
\\&
\left.
\hspace{1in}\times \exp(-c x_2^2/(4\beq^2)-(1-c) x_1^2/(4\beq^2)) 
\dd x_1\dd x_2\right]\label{eq:sfpRIND}.
\end{align}

Consider the special case $\sbangle=\pi/2$. Here $c=1/2$ and the last term can be further solved owing to the symmetry of the inner terms with respect to $x_1$ and $x_2$:
\begin{align}
\pRIND=&c\left[2-\frac{2\beq}{1-c}\left({\Q(\beq)}-\frac{\Q(\beq/\sqrt{c})}{\sqrt{c}}\right)-\frac{c_2\beq}{2(1-c)^2}{\left(1-\frac2{\sqrt{1-c}}\Q\left(\frac{\beq}{\sqrt{1-c}}\right)\right)}^2\right]\label{eq:sfpRIND}.
\end{align}

The lower bounds computed in the previous subsections can similarly be obtained for the self-blocking case by replacing the joint distribution of $R_1,R_2$ with the joint distribution of $R_1$, $D_2$ and $\Phi$ and adding the probability of self-blocking in terms $\prob{\event{A}_1},\prob{\event{A}_2}$ and $\prob{\event{A}_1\cup\event{A}_2}$. 
The asymptotic lower bound is given as
\begin{align}
\pRDEP&=c\left(2-\frac{2\beq}{1-c}\left({\Q(\beq)}-\frac{\Q(\beq/\sqrt{c})}{\sqrt{c}}\right)\right)\nonumber\\
&
-\frac{c_2}{2\beq^2\pi}\int_0^\infty\int_\sbangle^\pi S(x)
\exp(-x\sin^2(\Phi/2)-(1-c)x^2/4\beq^2) x\dd x
\end{align}
where
\begin{align}
&S(x)=\frac{1}{c}
\exp\left(\beq_c^2-\left(
    \frac{1}{2\beq_c} x  +\beq_c
    \right)^2\right)
    \left(1-2\beq_c \Q\left(
    \frac{1}{2\beq_c} x  +\beq_c
    \right)\right).
\end{align}

%
%
%
%
%
%
%



\section{Reliability Analysis for $n\ths$ Order Macro Diversity}
We can extend the analysis performed for $n=2$  case to the general $n$ case. In this section, we consider the general case with $n\ths$ order diversity.  In this case, the typical user at the origin $\point{O}$ is connected with $n$ BSs $\point{B}_1\cdots\point{B}_n$ with link lengths equal to $R_1\cdots R_n$. Let us denote the angles between  the link $\point{\Z}_n$ and other links $\point{\Z}_1, \point{\Z}_2\cdots \point{\Z}_{n-1}$ respectively as $\Phi_1\cdots \Phi_{n-1}$. Without loss of generality, we assume that $\mathbf{x}_n$ is at $x$ axis. The joint distribution of $R_n$'s is given as (see Appendix \ref{appen:njointdist}):
\begin{align}
f(r_1,r_2,\cdots,r_n)&=(2\pi\lambda)^nr_1r_2\cdots r_n\exp(-\lambda\pi r_n^2), \text{ if }r_1\le r_2\le \cdots \le r_n\label{eq:njointdist}
\end{align}
and $\Phi_i\sim \text{Uniform}(0,2\pi)$. Following the similar arguments as taken in the second order case, the reliability can be computed which is given in the following Theorem. 

\begin{figure}[ht!]
\begin{center}
\includegraphics[width=2.5in]{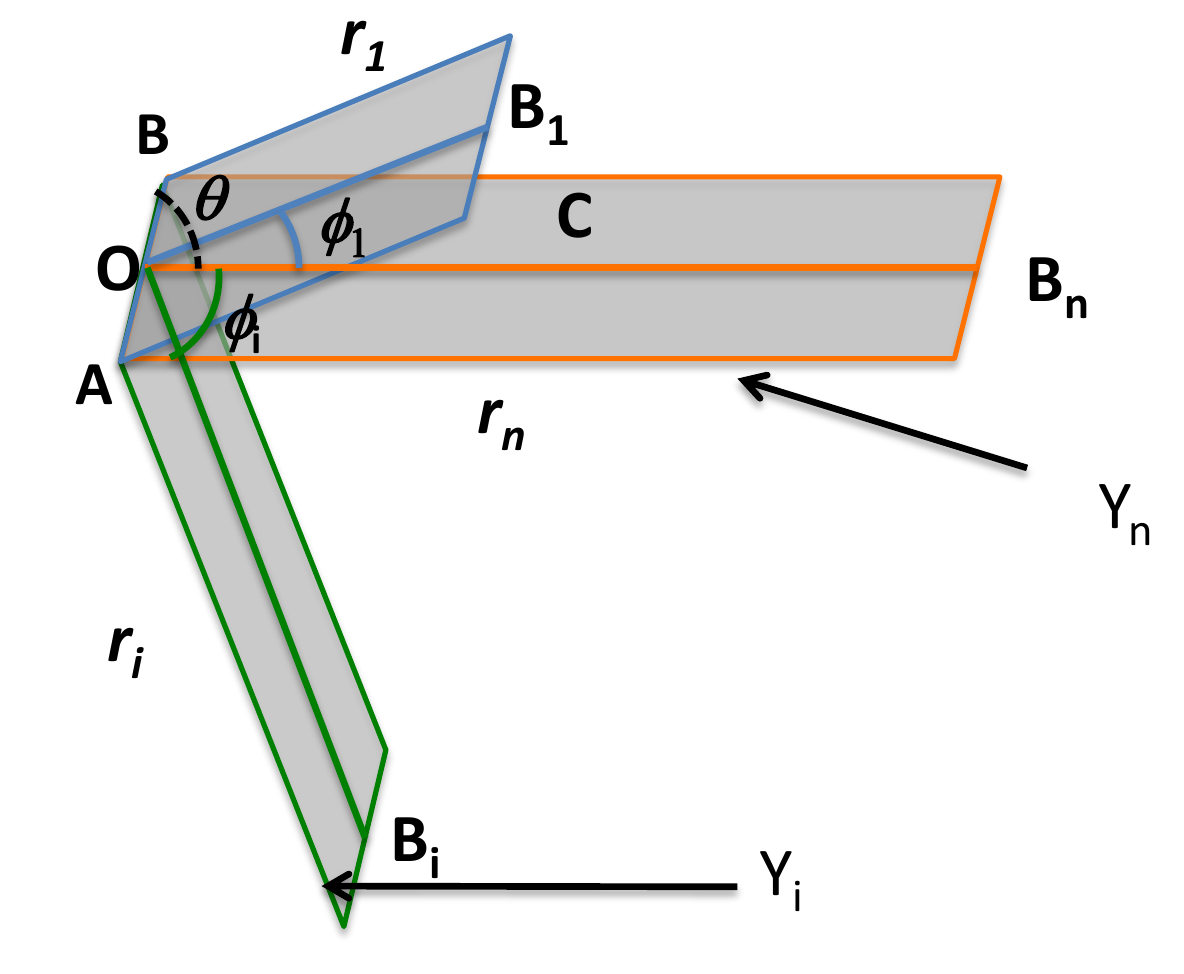}
\caption{The union of $n$ parallelograms $\mathcal{P}_i$'s described in Theorem \ref{thm:nDEP}.}
\label{fig:blockingarean}
\end{center}
\end{figure}

\begin{theorem}\label{thm:nDEP}
Let $\mathcal{P}_i$ denote a parallelogram  with sides $\point{Y}_i$ (with length $r_i$  and orientation $\phi_i$) and $\point{AB}$ (with length $\ell$ and orientation $\theta$) as shown in Fig. \ref{fig:blockingarean}.
Let $A(S,\{r_i\},\{\phi_i\},\ell,\theta)$ is the area of union of parallelograms $\mathcal{P}_i$'s ($i\in S$) where $S$ is a subset of $\{1,2,\cdots,n\}$. Let $\s(\{r_i\},\{\phi_i\})$ denote the average of the area $A$ over $(L,\Theta)$ which is given as
\begin{align}
\s(S,\{r_i\},\{\phi_i\})&=\int_0^\infty \int_0^\pi  A(S,\{r_i\},\{\phi_i\},\ell,\theta)F_L(\dd\ell)F_\Theta(\dd\theta).
\end{align}
Now, the reliability probability in a cellular network with $n\ths$ order of macro diversity is 
\begin{align*}
\pR&=\frac{1}{(2\pi)^{n-1}}\int_{[0,2\pi]^{n-1}}\int_{{(\mathbb{R}^+)}^n} K(\{r_i\},\{\phi_i\}) f_{\{R_i\}}(\{r_i\}) \dd r_1 \dd r_2 \cdots \dd r_{n}\dd\phi_1 \cdots \dd \phi_{n-1}
\end{align*}
where  $f_{\{R_i\}}(\{r_i\})$ is the joint distribution of $R_i$'s given in \eqref{eq:njointdist} and
\begin{align*}
K(\{r_i\},\{\phi_i\})=\sum_{S:S\subset[1,n]}^n (-1)^{|S|-1}  \exp(-\lbuild \s(S,\{r_i\},\{\phi_i\})).
\end{align*}  
\end{theorem}


Due to large numbers of variables in Theorem \ref{thm:nDEP}, it is not possible to analytically solve the expression. Hence, we consider the two special cases to bound the reliability.

\subsection{Independent Blocking}
We first consider the independent blocking case. As argued in the $n=2$ case, the reliability in the independent blocking scenario provides an upper bound to the reliability in the dependent blocking scenario. 

Let $\event{A}_j$ denote the probability that $j\ths$ link ($\point{Z}_j$) is not blocked. Hence, the probability that at least one link is unblocked is given as
\begin{align*}
\prob{\cup \event{A}_j}&=1-\prob{\cap \event{A}_j^\complement}=1-\prod_{j=1}{n}\prob{\event{A}_j}
=1-\prod_{j=1}^{n}(1-\exp(-\beta R_j)).
\end{align*}
Therefore the reliability is given as
\begin{align}
\pRIND&=1-\int_{0}^{\infty} \int_{0}^{\infty}\cdots\int_0^{\infty} (1-e^{-\beta r_1}) (1-e^{-\beta r_2}) \cdots(1-e^{-\beta r_n})  f(r_1,r_2,\cdots,r_n)  dr_1 dr_2\cdots r_n\nonumber \\
&=1-(2\pi\lambda)^n\int_{0}^{\infty} \int_{0}^{r_n}\cdots\int_0^{r_2} (1-e^{-\beta r_1}) (1-e^{-\beta r_2}) \cdots(1-e^{-\beta r_n})  r_1r_2r_n\exp(-\lambda\pi r_n^2)  dr_1 dr_2\cdots r_n \nonumber\\
&=1-\frac{2 (2\beq)^{-2n-2}}{\Gamma(n+1)}  \int_0^\infty e^{-t^2/(4\beq^2)}
\left[t^2-2+2e^{-t}(t+1)\right]^ndt\label{eq:npRIND}
\end{align}
where the last step is due to mathematical induction. The complete proof can be found in  Appendix \ref{appen:npRIND}. 

\subsection{Dependent Blocking in The Presence of Uniform Blockages}
We now consider a cellular system with blockage process where blockage lengths and orientations are  uniformly distributed. Due to the number of variables, the area of the union of parallelograms $\mathcal{P}_j$ is difficult to evaluate. Hence, we provide a tractable lower bound for the dependent blockage case.
 
 Let $A_j$ denote the area of the parallelogram $\mathcal{P}_j$. Now consider a set $S=\{i_1,i_2,\cdots,i_k\}\subset\{1,2,\cdots,n\}$ with increasing order of indexes. Without loss of generality assume that  $\Phi_{i_k}=0$. Given $\theta$, let $\mathcal{E}_j$ denote event that $(\theta\le\Phi_{i_j}\le \pi+\theta)$. This event is equivalent to the condition that the parallelogram $\mathcal{P}_{i_j}$ does not overlap with parallelogram $\mathcal{P}_{i_k}$. It can be seen that there can not be two or more mutually disjoint parallelograms which  do not overlap with $\mathcal{P}_{i_k}$.
 
  Now, we can bound the Area $A(S,\ell,\theta)$ from below as follows:
\begin{align}
A(S,\ell,\theta)&\ge A_{i_k}
+A_{i_{k-1}} \indside{\mathcal{E}_{i_{k-1}}}
+A_{i_{k-2}} \indside{\mathcal{E}_{i_{k-1}}^\complement\cap\mathcal{E}_{i_{k-2}}}
+
A_{i_{k-3}} \indside{\cap_{j=1}^2\mathcal{E}_{i_{k-j}}^\complement\cap\mathcal{E}_{i_{k-3}}}\nonumber\\&
+\cdots+
A_{i_{k-l}} \indside{\cap_{j=1}^{l-1}\mathcal{E}_{i_{k-j}}^\complement\cap\mathcal{E}_{i_{k-l}}}+\cdots+
0 \cdot \indside{\cap_{j=1}^{k-1}\mathcal{E}_{i_{k-j}}^\complement}\label{eq:nALowerBound}.
\end{align}
In the  lower bound in \eqref{eq:nALowerBound}, we always include the area of the largest parallelogram $\mathcal{P}_{i_k}$. Now, if the next largest parallelogram $\mathcal{P}_{i_{k-1}}$ is not overlapping with $\mathcal{P}_{i_k}$ (which is equivalent to $\event{E}_{i_{k-1}}$), then we will include $\mathcal{P}_{i_k}$ in the lower bound. Now, as discussed above, there cannot be any other parallelogram $\mathcal{P}_{i_j}$ ($j\le k-1$) which does not overlap with either of the two parallelograms $\mathcal{P}_{i_k}$ and $\mathcal{P}_{i_{k-1}}$. But, if $\mathcal{P}_{i_{k-1}}$  overlaps with $\mathcal{P}_{i_k}$, then we will consider the next largest parallelogram $\mathcal{P}_{i_{k-2}}$. We continue the search until we get the one parallelogram disjoint to 
$\mathcal{P}_{i_k}$. If there are no such disjoint parallelograms, then we will keep only the area of $\mathcal{P}_{i_{k}}$ in the lower bound. 

 Now, integrating \eqref{eq:nALowerBound} with respect to $\theta$ and $\ell$ (see Appendix \ref{appen:npRLowerBound} for the the proof sketch) gives the following lower bound (denoted by $\LBII{\s}$) of function $\s(S,\{R_i\},\{ \Phi_i\})$:
\begin{align*}
\s(S,\{R_i\},\{\Phi_i\})\ge \LBII{\s}(S,\{R_i\},\{\Phi_i\})=\frac{\Lmax}{\pi} \left(r_{i_k}+\sum_{j=1}^{k-1} r_{i_j}\frac{1}{2^{j-1}}\sin^2(\Phi_{i_j}/2)\right).
\end{align*}

Now using  Theorem \ref{thm:nDEP}, a lower bound on the reliability can be computed:
\begin{align}
\pR\ge\pRDEP=&\frac{1}{(2\pi)^{n-1}}\frac{1}{2^n\beq^{2n}}\int_{[0,2\pi]^{n-1}}\int_{x_1\le x_2\cdots \le x_n} K(\{x_i\},\{\phi_i\}) \exp\left(-x_n^2/(4\beq^2)\right)\nonumber\\
&\times x_1x_2\cdots x_n \dd x_1 \dd x_2 \cdots \dd x_{n}\dd\phi_1 \cdots \dd \phi_{n-1}\label{eq:npRDEP}
\end{align}
where
\begin{align*}
K(\{x_i\},\{\phi_i\})=\sum_{k=1}^{n}\sum_{S:S\subset[1,n],|S|=k}^n (-1)^{k-1}  \exp\left(-x_{i_k}-\sum_{j=1}^{k-1} x_{i_j}\frac{1}{2^{j-1}}\sin^2(\phi_{i_j}/2)\right).
\end{align*}  

Similar to the $n=2$ case,  it can be shown that $\mu A(S,\ell,\theta)$ and hence reliability decreases with increasing $\Lmax$ for a given $\beta$, and the lower bound becomes asymptotic tight for large $\Lmax$ as $\Lmax\rightarrow\infty$. Now by combining the upper bound computed from the independent blocking, lower bound computed above and monotonicity of reliability, we get the following Theorem.
\begin{theorem}[General $n$ case]
The reliability of a cellular system with $n\ths$ order macro diversity in presence of a blockage process (with $\ell\sim U(0,\Lmax)$ and $\theta\sim U(0,\pi)$) with fixed $\beta$, is bounded as
\begin{align}\pRDEP(\beta/\sqrt{\lbs})\le\pR(\lbs,\lbuild,\Lmax)\le
\pRIND(\beta/\sqrt{\lbs})  
\end{align}
where $\pRIND$ and $\pRDEP$ is given in \eqref{eq:npRIND} and \eqref{eq:npRDEP}. The two bounds are achieved when the maximum blockage size $\Lmax$ is 0 and $\infty$ respectively.
\end{theorem}

\newcommand{\insertIFig}{\begin{figure}[ht!]
\begin{center}
\includegraphics[width=3.7in]{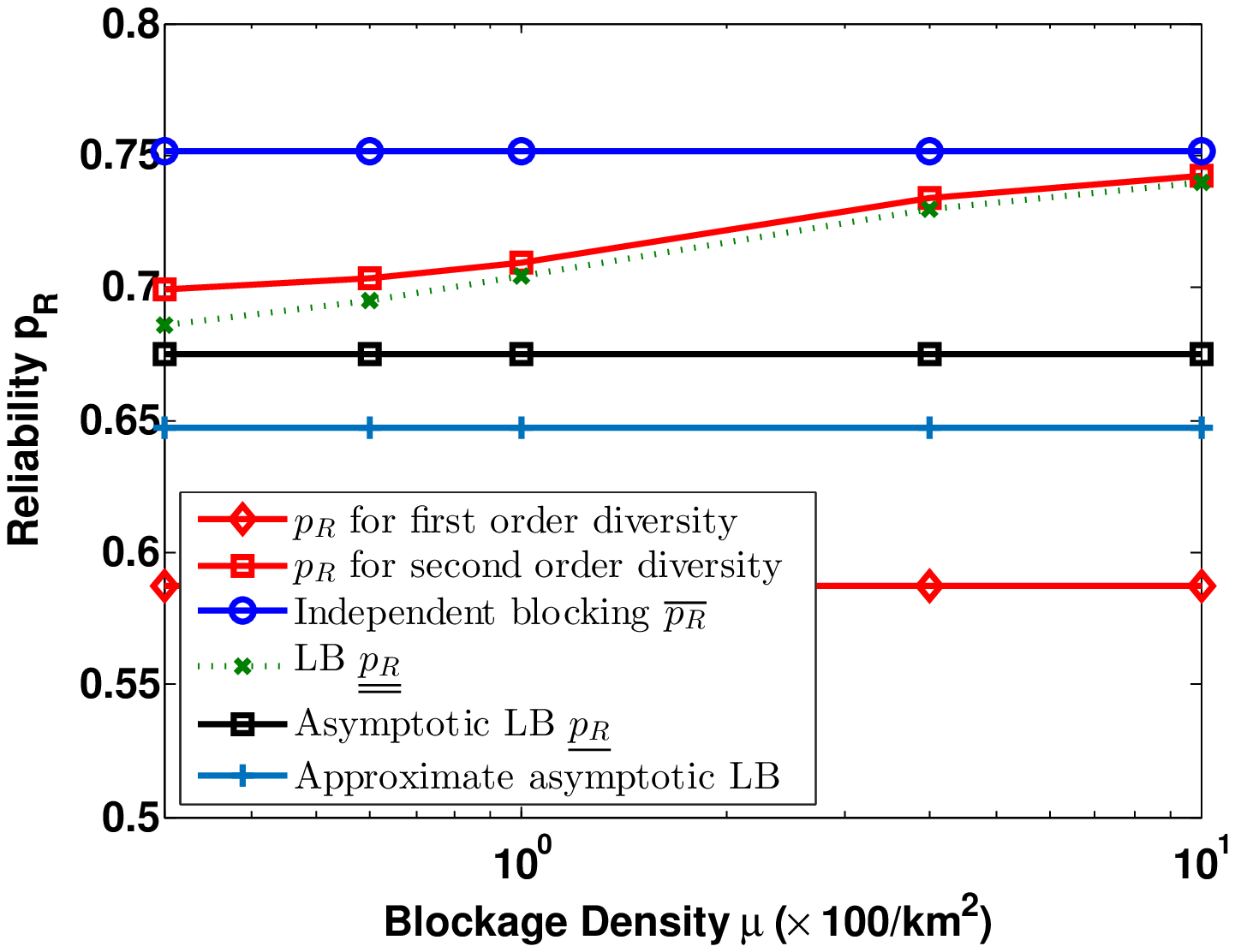}
\caption{Reliability in presence of blockages with fixed $\beta$ and varying density ($\lbuild$) in a cellular system with second order macro diversity. The $\pR$ for independent blockage case and computed lower bounds (LB) are also shown.}
\label{fig:fig1}
\end{center}
\end{figure}}

\newcommand{\insertIIFig}{
\begin{figure}[ht!]
\begin{center}
\includegraphics[width=3.7in]{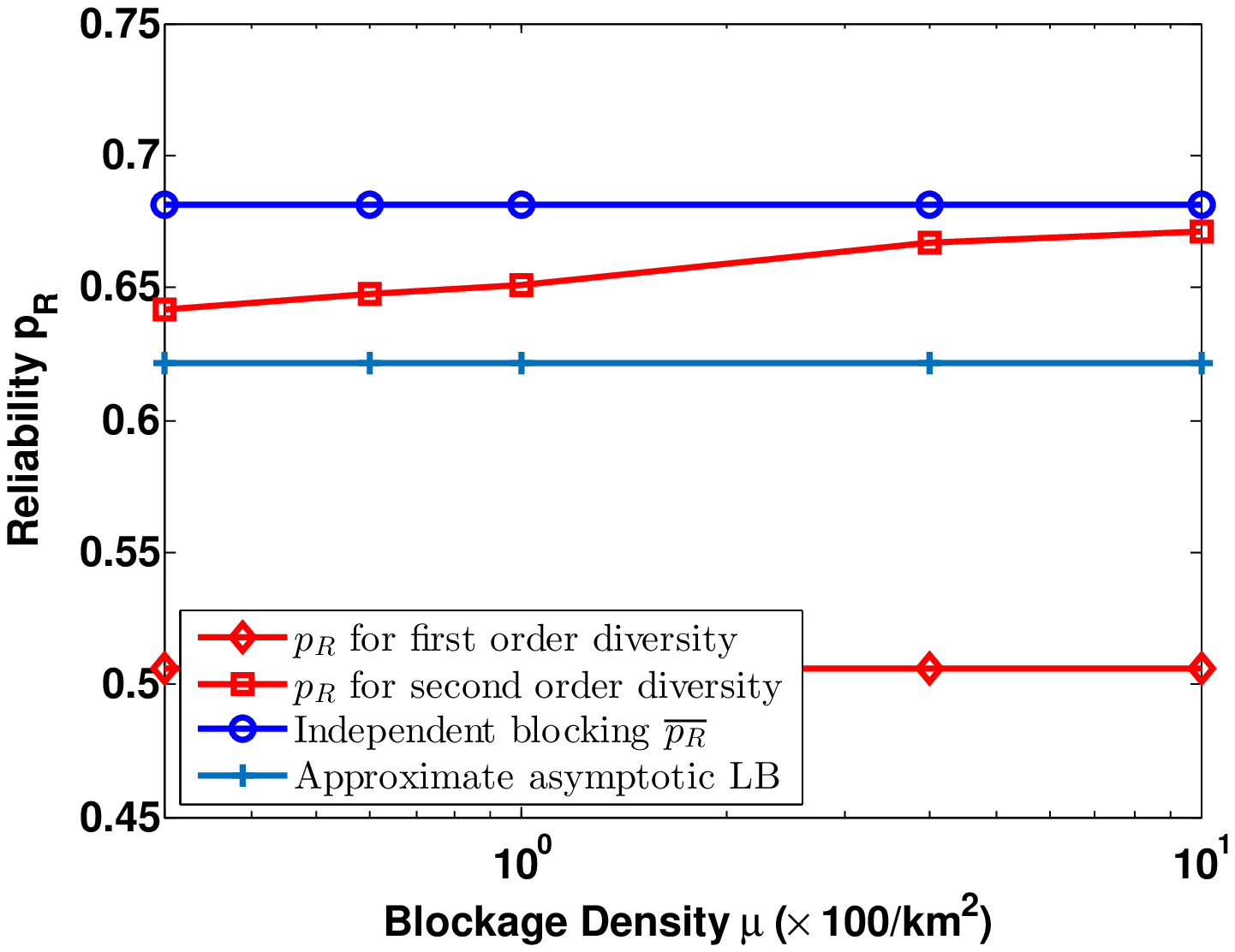}
\caption{Reliability in the presence of self-blocking and blockages with fixed $\beta$ and varying  density ($\lbuild$) in a cellular system with second order macro diversity. The $\pR$ for independent blockage case and computed lower bounds are also shown.}
\label{fig:fig2}
\end{center}
\end{figure}}

\newcommand{\insertIIIFig}{
\begin{figure}[ht!]
\begin{center}
\includegraphics[width=3.7in]{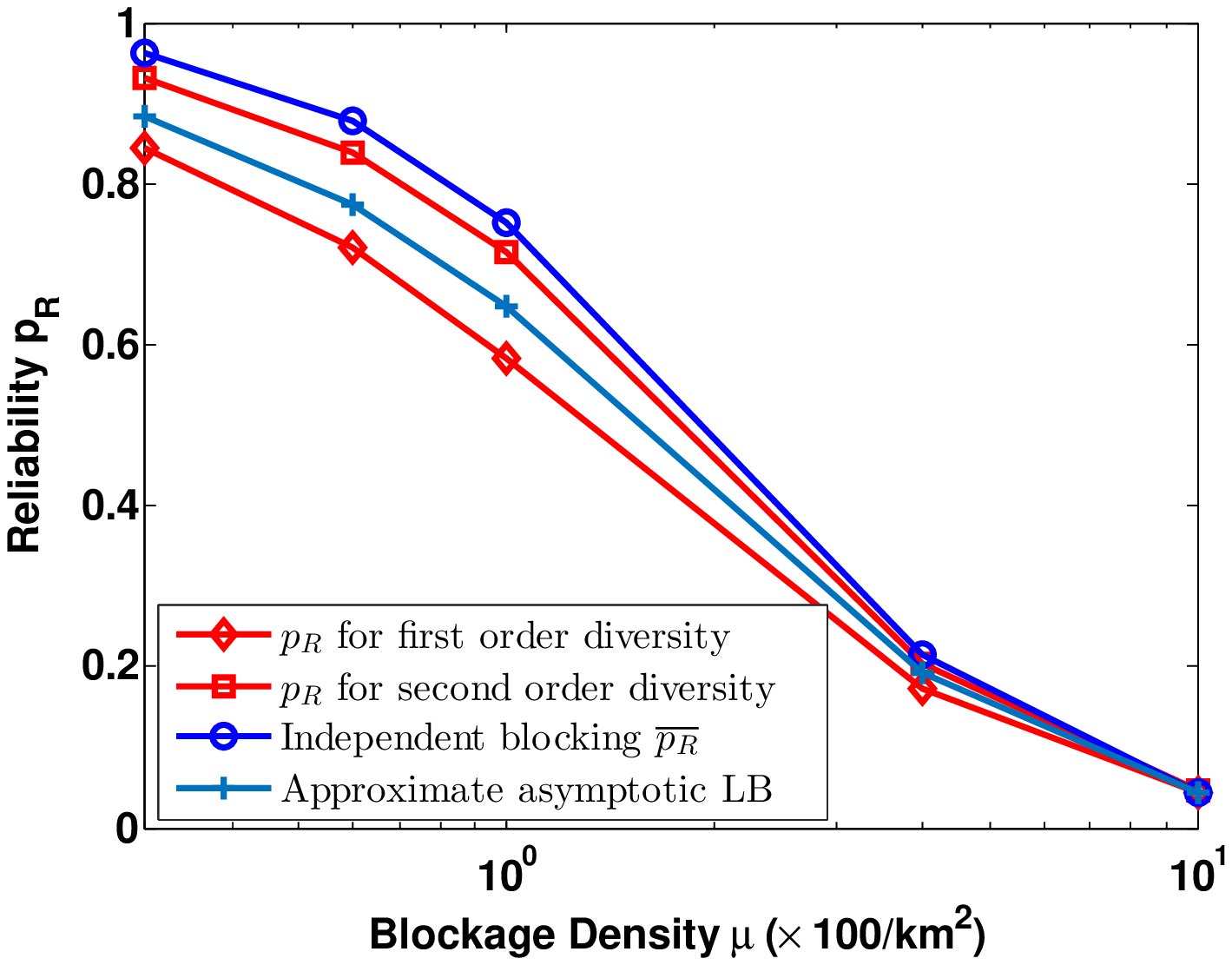}
\caption{Reliability in the presence of blockages with fixed $\Lmax=100m$ and varying density ($\lbuild$) in a cellular system with second order diversity. The $\pR$ for the independent blockage case and computed lower bounds are also shown. The bounds become more tight for larger blockage density.}
\label{fig:fig3}
\end{center}
\end{figure}}

\newcommand{\insertIVFig}{
\begin{figure}[ht!]
\begin{center}
\includegraphics[width=3.7in]{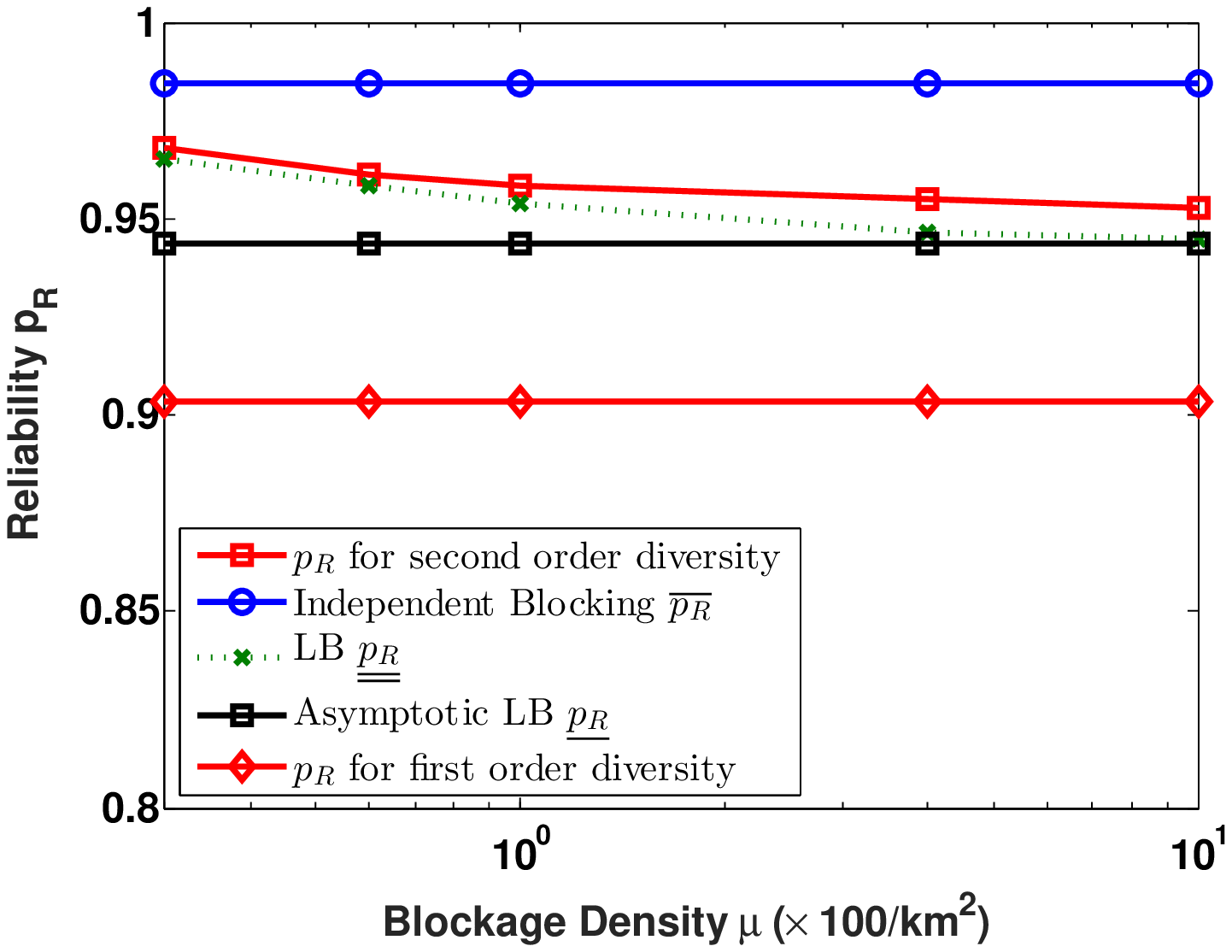}
\caption{Reliability in the presence of blockages with fixed $\Lmax=100m$ and varying blockage density ($\lbuild$) and scaling BS density ($\lbs$)  as $\lbuild^2$ in a cellular system with second order macro diversity. $\pR$ for the independent blockage case and computed lower bounds are also plotted. }
\label{fig:fig4}
\end{center}
\end{figure}}

\newcommand{\insertVIFig}{
\begin{figure}[ht!]
\begin{center}
\includegraphics[width=3.7in]{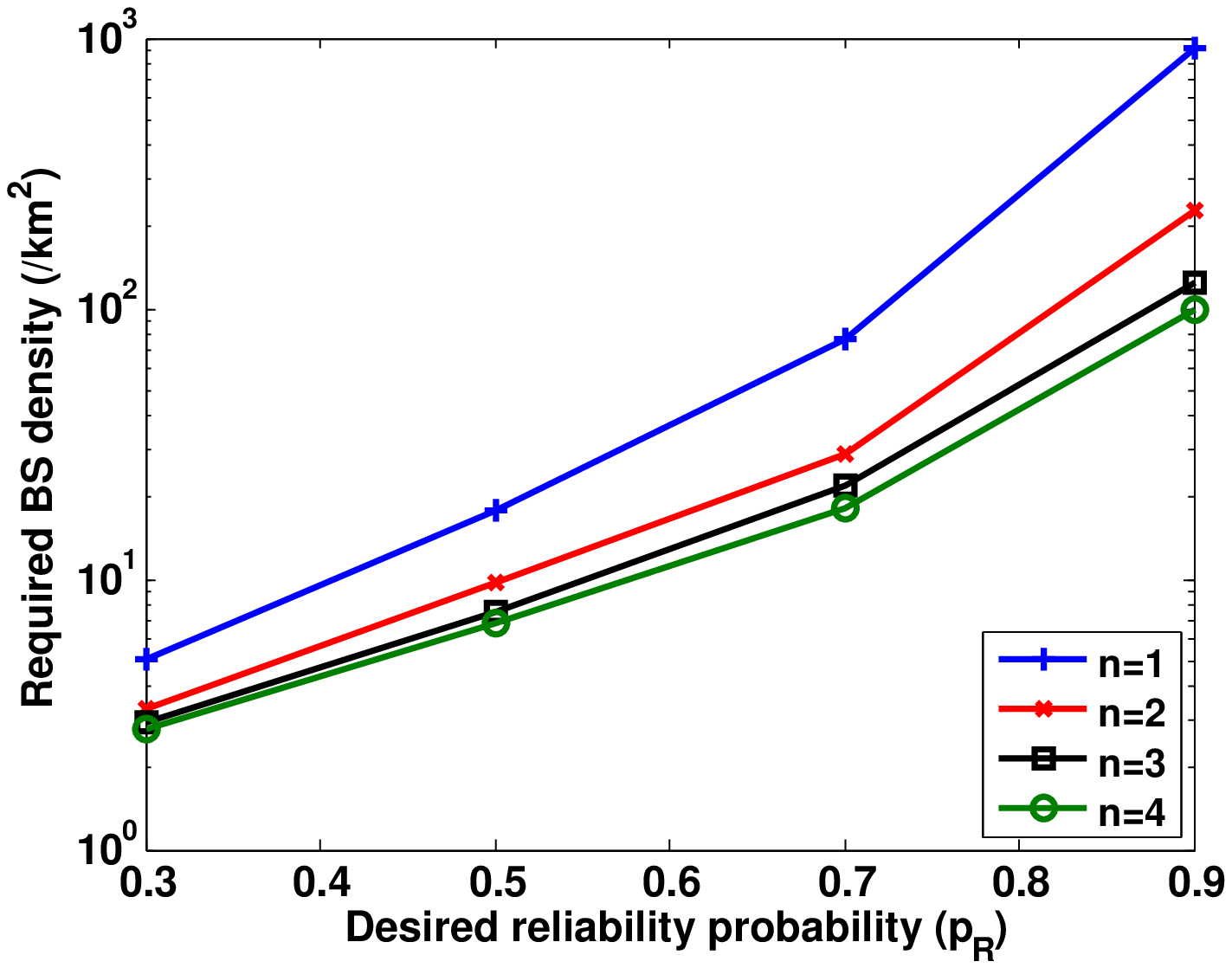}
\caption{Required BS density versus desired reliability for various macro diversity order ($n$) in the independent blockage case. Higher macro diversity order can reduce the required BS density by order of magnitudes.}
\label{fig:fig5}
\end{center}
\end{figure}}

\newcommand{\insertDataFig}{
\begin{figure}[ht!]
\begin{center}
\includegraphics[width=0.46\textwidth]{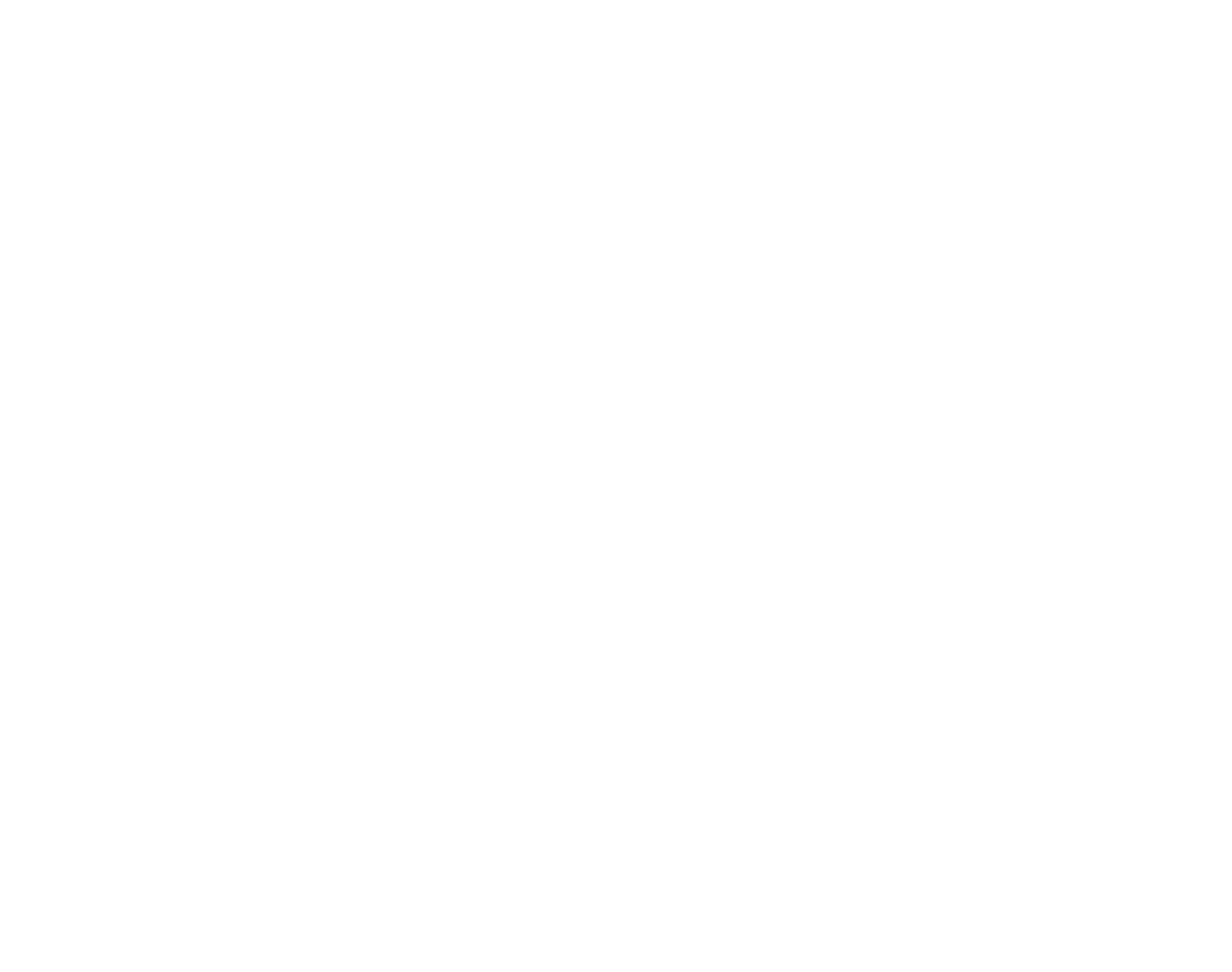}
\includegraphics[width=0.46\textwidth]{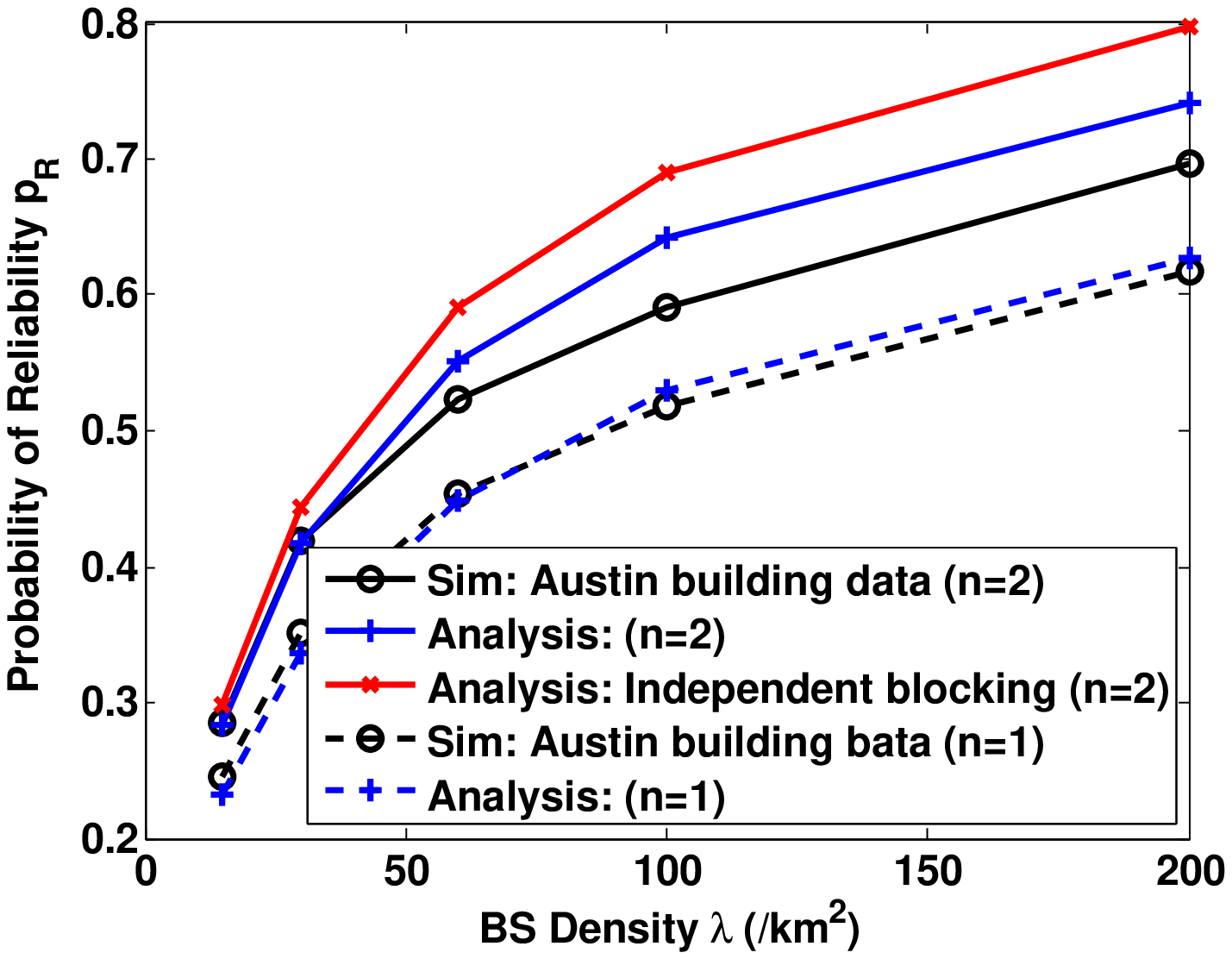}
\caption{Validation of analysis with real building data (a) The used building map near The University of Texas at Austin. The rectangular area in the center denotes the locations of generated users. BSs are uniformly generated over the whole space. (b) The reliability for a  cellular system with first and second order macro diversity.}
\label{fig:figdata}
\end{center}
\end{figure}}

\insertDataFig

\insertIFig
\section{Numerical Results}\label{sec:results}
In this section, we present  numerical results to evaluate the bounds and also to draw insights into the gains of macro diversity. 
We consider a blockage process with uniform distribution of blockages lengths and orientation and with parameter $\beta$. 
The BS density is assumed to be $\lbs=30$ BS/km$^2$ which corresponds to an average inter-site distance of $100$ m.

{\textbf{Validation with Real Building Data.}} To validate our analysis, we consider a region near The University of Texas at Austin \cite{SinghBackHaul2015} as shown in Fig. \ref{fig:figdata}(a) with BSs location modeled as PPP and users uniformly located in the smaller rectangle uniformly. For a system with second order macro diversity, we plot the actual reliability probability with the one computed from the analysis in Fig. \ref{fig:figdata}(b). The parameters are obtained by fitting the reliability $\pR$ for a single BS link ($n=1$) and are given as  $\beta=0.014/$m and $\lbuild=2.2\times10^{-4}$m$^2$. It can be observed that analysis approximates the performance in the real scenario quite well. 

\insertIIFig

{\textbf{Impact of Blockage Correlation.}} We now show the impact of blockage correlation by decreasing the blockage size with fixed $\beta$. Fig. \ref{fig:fig1} shows the variation of reliability 
with respect to blockage density $\lbuild$ while keeping $\beta$ fixed at $6.4$km$^{-1}$. When compared to first order diversity case (which means no diversity), the second diversity can increase the reliability probability by 35\%. As shown in analysis, the reliability decreases when $\Lmax$ increases or $\lbuild$ decreases. Fig. \ref{fig:fig1} also shows  $\pR$ for the independent blocking case and asymptotic lower bound for $\Lmax\rightarrow\infty$ case. It can be seen that $\pR$ reaches the independent blocking case for high blockage density and low blockage size. This result shows  that correlation in blockages can decrease the reliability probability  by 15\%.

\insertIIIFig

 {\textbf{Impact of Self-Blocking.}} We now consider a cellular system with second order macro diversity and self-blocking with a blocking angle of $60^{\omicron}$. Fig. \ref{fig:fig2} shows the variation of reliability 
 with respect to blockage density $\lbuild$ while keeping $\beta$ fixed at $6.4$km$^{-1}$. Due to self-blocking the reliability has further decreased than the case with no self-blocking. Fig. \ref{fig:fig2} also shows  $\pR$ for the independent blocking case and the asymptotic lower bound for the $\Lmax\rightarrow\infty$ case. It can be seen that $\pR$ decreases when the maximum blockage length $\Lmax$ increases  (the blockage density  $\lbuild$ decreases) and  reaches $\pRIND$ for high blockage density. 

\insertIVFig
 \insertVIFig
 
{\textbf{Impact of Blockage Density and Scaling.}} Fig. \ref{fig:fig3} shows the variation of reliability 
with respect to blockage density $\lbuild$  while keeping $\Lmax$ fixed at $100$m and with fixed BS density, along with  $\pR$ for the independent blocking case and asymptotic lower bound for $\Lmax\rightarrow\infty$ case. It can be obeserved that the bounds for $\pR$ become tighter for higher blockage density. Since the BS density is kept fixed,  the reliability decreases significantly with $\lbuild$.  To show the required scaling requirements, we show in Fig. \ref{fig:fig4}, variation of reliability 
with respect to blockage density $\lbuild$ while keeping $\Lmax$ fixed at $100$m and scaling BS density as $\lbuild^2$. It can be seen that  the reliability decreases slightly with $\lbuild$ but remains quite flat  with constant upper and lower bounds. This implies that BS density should scale as $\lbuild^2$ to keep the same level of LOS connectivity in the system.

{\textbf{Impact of Macro Diversity.}} We now show the gain of macro diversity. We assume uniform blockage with  density $\lbuild=100/$km$^2$ and maximum blockage length $\Lmax=100$m which is equivalent to $\beta =6.4$km$^{-1}$. 
Fig. \ref{fig:fig5} shows required density (obtained from solving the reverse problem) as a function of $\pR$ for various diversity order for the independent blocking case.  It can be seen that if each user can be connected to four BSs at any time, the required BS density to achieve a certain reliability is decreased by order of tens. In particular, for $\pR=0.9$, the required BS density for $n=4$ is 90 BS/km$^2$ which is 10 times less than the required BS density for $n=1$.

 \section{Conclusions}\label{Sec:Conclusions}
We have evaluated the gains of macro diversity for a mmWave cellular system in the presence of random blockages. We proposed a framework to analyze the correlation among blocking of multiple links in a cellular system and computed the system's reliability. We also study the impact of blockage sizes for linear blockages and show that correlation in blockages decreases the macro diversity gain. We also compared different uniform blockage processes while keeping the product of blockage density and blockages length fixed and showed that macro diversity gains are higher when blockage lengths are small. We also show that BS density should scale as square of blockage density to maintain  a certain level of system reliability.  
 The work has numerous possible extensions. First, the proposed framework can be extended to analyze the coverage probability and rate coverage in a system with multi-BS diversity. Second, the framework can used to develop a correlated shadowing model to study the impact of correlated shadowing on cellular systems' performance. Third, the framework can be extended to include multi-cell cooperation where a user is served simultaneously by multiple BSs. 

\appendices
\section{Proof of Lemma \ref{lemma:monotone}} \label{appen:prooflemmamonotone}
To prove the Lemma, we show that $\lbuild \s(R_1,R_2,\Phi)$ for the blockage with distribution $F'_c(\dd \ell)$ is less than  $\lbuild \s(R_1,R_2,\Phi)$ for the blockage with distribution $F_L(\dd \ell)$ for $c>1$.  For the rest of the proof, we assume $c>1$. It can been shown easily that given $\ell$ and $\theta$,
\begin{align*}
\frac{1}{c}A(R_1,R_2,\phi,c\ell ,\theta)\le A(R_1,R_2,\phi,\ell,\theta).
\end{align*}
 Now, for the blockage process with distribution $F'_c(\dd \ell)$, $\lbuild \s(R_1,R_2,\Phi)$ is given as
\begin{align*}
\lbuild \s_c(R_1,R_2,\Phi)&=\beta \pi \frac{ \int_0^\infty \int_0^\pi  A(R_1,R_2,\phi,\ell,\theta)F'_c(\dd\ell)F_\Theta(\dd\theta)}{\int_0^\infty \ell F'_c(\dd \ell)}\\
&=\beta \pi \frac{ \int_0^\infty \int_0^\pi  A(R_1,R_2,\phi,\ell,\theta)F_L(\dd\ell/c)F_\Theta(\dd\theta)}{\int_0^\infty \ell F_L(\dd \ell/c)}\\
&=\beta \pi \frac{ \int_0^\infty \int_0^\pi  A(R_1,R_2,\phi,c\ell',\theta)F_L(\dd\ell/c)F_\Theta(\dd\theta)}{c\int_0^\infty \ell' F_L(\dd \ell')}\\
&\le\beta \pi \frac{ \int_0^\infty \int_0^\pi  A(R_1,R_2,\phi,\ell',\theta)F_L(\dd\ell/c)F_\Theta(\dd\theta)}{\int_0^\infty \ell' F_L(\dd \ell')}=\lbuild \s(R_1,R_2,\Phi)
\end{align*}
which completes the proof.

\section{Proof of Theorem \ref{thm:n2IND} }\label{appen:thmn2IND}
For the independent blocking case, the reliability is given as
\begin{align*}
\pRIND&=\expects{R_1,R_2,\Phi}{\prob{\event{A}_1\cup\event{A}_2}}=\expects{R_1,R_2,\Phi}{1-\prob{\event{A}_1^\complement\cap\event{A}_2^\complement}}\\
&=1-\expects{R_1,R_2,\Phi}{\prob{\event{A}_1^\complement}\prob{\event{A}_2^\complement}}
\end{align*}
where the last step is due to independence of events $\event{A}_1^\complement$ and $\event{A}_2^\complement$. Now, using \eqref{eq:prexp1} and \eqref{eq:prexpB}, we get
\begin{align}
\pRIND&=1-\int_{0}^{\infty} \int_{0}^{\infty} (1-e^{-\beta r_1}) (1-e^{-\beta r_2}) f(r_1,r_2)  \dd r_1 \dd r_2 \label{eq:pRIND1}\\
&=1-(2\pi\lbs)^2\int_{0}^{\infty} (1-e^{-\beta r_2})r_2 e^{-\lambda\pi r_2^2}\int_{0}^{r_1} (1-e^{-\beta r_1})  r_1  \dd r_1 \dd r_2.
\end{align}
Now using the transformations $x_1=\beta r_1$ and $x_2=\beta r_2$, we get
\begin{align*}
\pRIND&=1-\frac{1}{4\beq^4}\int_{0}^{\infty} (1-e^{-x_2})x_2 e^{-x_2^2/(4\beq^2)}\int_{0}^{x_1} (1-e^{-x_1})  x_1  \dd x_1 \dd x_2\\
&=1-\frac{1}{8\beq^4} \int_0^\infty e^{-x_2^2/(4\beq^2)}x_2(1-e^{-x_2})[x_2^2-2+2e^{-x_2}(x_2+1)]\dd x_2\\
%
&=\frac{1}{\beq}\left[\beq^3-\Q(\beq)(2\beq^4+5\beq^2-1)+\Q(2\beq)(8\beq^2-1)\right].
\end{align*}
\section{Proof for Asymptotic Lower Bound }\label{appen:proofpRDEP2}
Using Theorem \ref{thm:n2DEP} and the lower bound of $\s(R_1,R_2,\Phi)$ derived in \eqref{eq:sLowerBound2}, the lower bound on the reliability probability can be given as:
\begin{align}
\pRDEP=&2+\beq^2-\beq(5+2\beq^2)\Q(\beq)-\nonumber\\
&\frac{1}{4\beq^4\pi}\int_{0}^{\pi}\int_{0}^{\infty} x_2\exp\left(-\frac{x_2^2}{4\beq^2}\right) \int_{0}^{x_2} \exp\left(-x_1\sin^2(\phi/2)-x_2\right) x_1 \dd x_1 \dd x_2 \dd\phi\nonumber\\
=&2+\beq^2-\beq(5+2\beq^2)\Q(\beq)-\nonumber\\
&\frac{1}{4\beq^4\pi}\int_{0}^{\pi}\int_{0}^{\infty} x_2\exp\left(-x_2-\frac{x_2^2}{4\beq^2}\right) \int_{0}^{x_2} \exp\left(-x_1\sin^2(\phi/2)\right) x_1 \dd x_1 \dd x_2 \dd\phi\nonumber\\
=&2+\beq^2-\beq(5+2\beq^2)\Q(\beq)-\nonumber\\
&\frac{1}{4\beq^4\pi}\int_{0}^{\pi}\int_{0}^{\infty} x_2\exp\left(-x_2-\frac{x_2^2}{4\beq^2}\right) 
\frac{1- \exp\left(-x_2\sin^2(\phi/2)\right)(1+x_2\sin^2(\phi/2)) }{\sin^2(\phi/2)}
     \dd x_2 \dd\phi\nonumber.
     \end{align}
     Now, using a $\phi/2\rightarrow\phi$ substitution and some manipulations, we get
     \begin{align}
=&2+\beq^2-\beq(5+2\beq^2)\Q(\beq)-\frac2\pi   
\frac{1}{4\beq^4}\int_{0}^{\pi/2}\frac{1}{{\sin^4(\phi)}}
\left[
\int_{0}^{\infty} x\exp(-x- x^2/(4\beq^2)) \dd x \right.\nonumber\\
&\left. -\sin^2(\phi)\int_{0}^{\infty} x^2\exp(-x(1+\sin^2(\phi))- x^2/(4\beq^2))  \dd x\right.\nonumber\\
&\left.-
 \int_{0}^{\infty} x\exp(-x(1+\sin^2(\phi))- x^2/(4\beq^2)) ) \dd x \right]\nonumber.
 \end{align}
 Now, by evaluating the inner integral, we get
 \begin{align}
\pRDEP= &2+\beq^2-\beq(5+2\beq^2)\Q(\beq)-\frac2\pi   
\frac{1}{4\beq^4}\int_{0}^{\pi/2}\frac{1}{{\sin^4(\phi)}}
\left[
2{\beq^2}(1-2\beq\Q(\beq))
\right.\nonumber\\
&\left. -\sin^2(\phi)
4\beq^3
\left(\Q((1+\sin^2(\phi))\beq) (2(1+\sin^2(\phi))^2\beq^2+1)-(1+\sin^2(\phi))\beq\right)\right.\nonumber\\
&\left.
- 2{\beq^2}(1-2\beq (1+\sin^2(\phi))\Q((1+\sin^2(\phi))\beq))
         \dd x \right]\nonumber.
 \end{align}

 Now, after some further manipulations, we get 
 \begin{align}
 \pRDEP=&1+\beq^2-\beq(5+2\beq^2)\Q(\beq)+\frac{4\beq}\pi   
\int_{0}^{\pi/2}
(2+\sin^2(\phi)) \Q((1+\sin^2(\phi))\beq))\dd\phi\nonumber
\\
&+\frac2\pi   
\frac{1}{\beq}\int_{0}^{\pi/2}
\left[
\Q(\beq)
- \Q((1+\sin^2(\phi))\beq))\right.\nonumber\\
&\hspace{1in}\left.+2\beq^2\sin^2(\phi)\Q((1+\sin^2(\phi))\beq))
-\sin^2(\phi)\beq\right]{\mathrm{cosec}^4(\phi)}\dd\phi\nonumber.
\end{align}

\section{Proof for the Approximate Linear Asymptotic Lower Bound} \label{appen:proofpRDEPlinearapprox}
We start the proof by noting that $\LBII{\s}$ in \eqref{eq:sLowerBound2} can be approximated as 
\begin{align}
s_\DEP(R_1,R_2,\Phi)&\approx \frac{\Lmax}{\pi}\left(R_2+R_1\frac{\Phi}{\pi}\right)\label{eq:sLinearApprox}.
\end{align}
Now, using Theorem \ref{thm:n2DEP} and \eqref{eq:sLinearApprox}, the lower bound on the reliability probability is given as:
\begin{align}
\pRDEP\approx&2+\beq^2-\beq(5+2\beq^2)\Q(\beq)-\nonumber\\
&\frac{1}{4\beq^4\pi}\int_{0}^{\pi}\int_{0}^{\infty} x_2\exp\left(-\frac{x_2^2}{4\beq^2}\right) \int_{0}^{x_2} \exp\left(-x_1\phi/\pi-x_2\right) x_1 \dd x_1 \dd x_2 \dd\phi\nonumber.
\end{align}
Now, by interchanging the limits for $x_1$ and $\phi$, we get
\begin{align}
\pRDEP\approx&2+\beq^2-\beq(5+2\beq^2)\Q(\beq)-\nonumber\\
&\frac{1}{4\beq^4\pi}\int_{0}^{\infty} x_2\exp\left(-x_2-\frac{x_2^2}{4\beq^2}\right) \int_{0}^{x_2} x_1 \int_{0}^{\pi}\exp\left(-x_1\phi/\pi\right) \dd\phi \dd x_1 \dd x_2\nonumber\\
=&2+\beq^2-\beq(5+2\beq^2)\Q(\beq)\nonumber\\
&-\frac{1}{4\beq^4\pi}\int_{0}^{\infty} \exp(-x_2-x_2^2/(4\beq^2))x_2 \int_0^{x_2}  x_1\frac{1- \exp(-x_1)}{x_1/\pi} \dd x_1 \dd x_2 \nonumber \\
=&2+\beq^2-\beq(5+2\beq^2)\Q(\beq)-\frac{1}{4\beq^4}\int_{0}^{\infty} \exp(-x_2-x_2^2/(4\beq^2))x_2 (x_2-1+ \expU{-x_2}) \dd x_2\nonumber\\
=&\frac1\beq\left[3\beq+\beq^3
-(7\beq^2+2\beq^4+2)\Q(\beq)+2\Q(2\beq)\right]\nonumber.
\end{align}

%
%

\section{Joint Distribution of $R_n$} \label{appen:njointdist}
For any $i\le n$, conditioned on the event $R_j=r_j (j\le i)$, the distribution of $R_{i+1}$ is given as
\begin{align}
\prob{R_{i+1}\le r_{i+1}|R_j=r_j,j\le i}&=\prob{\text{There exists at least one point in the ring } r_i\le r \le r_{i+1}}\nonumber\\
&=1-\exp\left(-\lbs\pi (r_{i+1}^2-r_i^2)\right).
\end{align}
Hence, the conditional PDF of $R_{i+1}$ is given as 
\begin{align}
f_{R_{i+1}}(r_{i+1}|R_j=r_j,j\le i)&=2\lbs\pi r_{i+1}\exp\left(-\lbs\pi (r_{i+1}^2-r_i^2)\right)\\
\implies f_{\{R_j\},R_{i+1}}(\{r_j\},r_{i+1})&=2\lbs\pi r_{i+1}\exp\left(-\lbs\pi (r_{i+1}^2-r_i^2)\right)f_{\{R_j\}}(\{r_j\})\label{eq:ithjointiterate}.
\end{align}
Now iterating \eqref{eq:ithjointiterate} for $n-1$ times from $i=n-1$ up to $i=1$, we get the joint distribution as follows
\begin{align}
f_{\{R_j\},j\le n}(\{r_j\})&=(2\lbs\pi)^n r_{n}r_{n-1}\cdots r_1 \prod_{i=1}^{n-1}\exp\left(-\lbs\pi (r_{i+1}^2-r_i^2)\right)\nonumber\\
&=(2\lbs\pi)^n r_{n}r_{n-1}\cdots r_1 \exp\left(-\lbs\pi r_{n}^2\right).
\end{align}

\section{Proof for  Independence Blocking Case with $n\ths$ Diversity Order}\label{appen:npRIND}
The reliability for the independence blocking case is given as
\begin{align*}
\pRIND&=1-\int_{0}^{\infty} \int_{0}^{\infty}\cdots\int_0^{\infty} (1-e^{-\beta r_1}) (1-e^{-\beta r_2}) \cdots(1-e^{-\beta r_n})  f(r_1,r_2,\cdots,r_n)  \dd r_1 \dd r_2\cdots \dd r_n \\
&=1-(2\pi\lambda)^n\int_{0}^{\infty} \int_{0}^{r_n}\cdots\int_0^{r_2} \prod_{i=1}^{n}r_i(1-e^{-\beta r_i})\exp(-\lambda\pi r_n^2)  \dd r_1 \dd r_2\cdots \dd r_n. 
\end{align*}
Using the substitutions $x_i=\beta r_i$, we get
\begin{align}
\pRIND&=1-\frac{(2\pi\lambda)^n}{\beta^{2n}}\int_{0}^{\infty} \int_{0}^{x_n}\cdots\int_0^{x_2} \prod_{i=1}^n x_i (1-e^{-x_i}) \exp\left(-\frac{\lambda\pi}{\beta^2} x_n^2\right)  \dd x_1 \dd x_2\cdots \dd x_n. \label{eq:ncaseproof1}
\end{align}
Let us define the function $J(n-i,y)$ by the following recursion 
\begin{align*}
J(i,y)&=\int_0^y t (1-e^{-t})J(i-1,t)\dd t\\
J(0,y)&=1.
\end{align*}
Then, \eqref{eq:ncaseproof1} can be written as
\begin{align}
\pRIND&=1-\frac{(2\pi\lambda)^n}{\beta^{2n}}\int_{0}^{\infty} \exp\left(- t^2/(4\beq^2)\right)  t (1-e^{-t}) J(n-1,t) \dd t  \label{eq:ncaseproof2}.
\end{align}
Now, we will prove the following using mathematical induction.
\begin{align}
J(i,y)=\frac{1}{2^i}\frac{1}{i!}\left[y^2-2+2(y+1)e^{-y}\right]^i.\label{eq:ncaseproofmI}
\end{align}
Step 1: For $i=0$,
\begin{align*}
J(0,y)=\frac{1}{2^0}\frac{1}{0!}\left[y^2-2+2(y+1)e^{-y}\right]^0=1.
\end{align*}
Step 2: Let us assume 
\begin{align*}
J(i,y)=\frac{1}{2^i}\frac{1}{i!}\left[y^2-2+2(y+1)e^{-y}\right]^i.
\end{align*}
Then
\begin{align*}
J(i+1,y)&=\int_0^y t (1-e^{-t})J(i,t)\dd t=\frac{1}{2^i}\frac{1}{i!}\int_0^y t (1-e^{-t})\left[t^2-2+2(t+1)e^{-t}\right]^i\dd t\\
&=\stackrel{(a)}{=}\frac{1}{2^{i+1}}\frac{1}{i!}\int_0^{y^2-2+2(y+1)e^{-y}}u^i\dd u\\
&=\frac{1}{2^{i+1}}\frac{1}{(i+1)!}\left[y^2-2+2(y+1)e^{-y}\right]^{i+1}
\end{align*}
which proves the identity \eqref{eq:ncaseproofmI}.
Using this identity in \eqref{eq:ncaseproof2}, we get
\begin{align*}
\pRIND&=1-\frac{2^{-2n+1}\beq^{-2n}}{(n-1)!}
\int_{0}^{\infty} \exp\left(- t^2/(4\beq^2)\right)  t (1-e^{-t})\left[t^2-2+2(t+1)e^{-t}\right]^{(n-1)} \dd t\\
&=1-\frac{2\cdot (2\beq)^{-2n-2}}{\Gamma(n+1)}  \int_0^\infty e^{-t^2/(4\beq^2)}
\left[t^2-2+2e^{-t}(t+1)\right]^ndt
\end{align*}
where the last step is due to integration by part.

\section{Proof Sketch for the Lower Bound for $n\ths$ Macro Diversity}\label{appen:npRLowerBound}
  Recall that given $\theta$, let $\mathcal{E}_j$ denote event that $(\theta\le\Phi_{i_j}\le \pi+\theta)$. Hence,
  \begin{align*}
  \prob{\mathcal{E}_j}=\expects{\theta}{\theta\le\Phi_{i_j}\le \pi+\theta}=\frac12.
  \end{align*}
Also, the area of $\mathcal{P}_{i_j}$ is given as
 \begin{align*}
 A_{i_j}&=R_{i_j}\ell\sin(|\theta-\Phi_{i_j}|).
 \end{align*}
 
Now, taking expectation of the both sides of \eqref{eq:nALowerBound} with respect to $\ell$ and $\theta$, we get 
\begin{align*}
&\expect{A(S,\ell,\theta)}\ge\expect{ A_{i_k}}+
\sum_{m=1}^{k} \expect{A_{i_{k-m}} \indside{\cap_{j=1}^{m-1}\mathcal{E}_{i_{k-j}}^\complement\cap
\mathcal{E}_{i_{k-m}}}}\\
&= R_{i_k}\expect{\ell} \ \expect{\sin(|\theta-\Phi_{i_k}|)}+
\sum_{m=1}^{k} R_{i_{k-m}}\expect{\ell}\ \expect{\sin(|\theta-\Phi_{i_{k-m}}|) \indside{\cap_{j=1}^{m-1}\mathcal{E}_{i_{k-j}}^\complement\cap
\mathcal{E}_{i_{k-m}}}}.
\end{align*}
Now $\event{E}_j$'s are mutually independent and also independent to $\Phi_m,j\ne m$. Therefore, we get
\begin{align*}
&\expect{A(S,\ell,\theta)}\ge R_{i_k}\frac{\Lmax}{\pi}+
\frac{\Lmax}{2}\sum_{m=1}^{k} R_{i_{k-m}}\expect{\sin(|\theta-\Phi_{i_{k-m}}|) \indside{
\mathcal{E}_{i_{k-m}}}}\prod_{j=1}^{m-1}\prob{\mathcal{E}_{i_{k-j}}^\complement}\\
 &=\frac{\Lmax}{\pi} \left(R_{i_k}+\sum_{j=1}^{k-1} R_{i_j}\frac{1}{2^{j-1}}\sin^2(\Phi_{i_j}/2)\right).
\end{align*}
Here, the last step is due to the following:
\begin{align*}
\expect{\sin(|\theta-\Phi_{i_{j}}|) \indside{
\mathcal{E}_{i_{j}}}}&=\frac1\pi\int_0^{\Phi_{i_j}} \sin(\Phi_{i_{j}}-\theta)\dd \theta=\frac1\pi\cos(\Phi_{i_j}-\theta)\big|_0^{\Phi_{i_j}}=\frac2\pi\sin^2(\Phi_{i_j}/2).
\end{align*}
%
%

\bibliographystyle{IEEEtran}
\bibliography{mmwavebib}

\end{document}